\documentclass[review]{elsarticle}

\usepackage{graphicx}
\usepackage{standalone}
\usepackage{amsmath}
\usepackage{float}
\usepackage{subcaption}
\usepackage{tkz-euclide}
\usepackage{algorithm, algpseudocode}

\usepackage{pgfplots}
\pgfplotsset{compat=1.13}
\usepgfplotslibrary{fillbetween}

\pgfmathdeclarefunction{gauss}{3}{%
	\pgfmathparse{1/(#3*sqrt(2*pi))*exp(-((#1-#2)^2)/(2*#3^2))}%
}

\definecolor{c0}{RGB}{228,26,28}
\definecolor{c1}{RGB}{55,126,184}
\definecolor{c2}{RGB}{77,175,74}
\definecolor{c3}{RGB}{152,78,163}
\definecolor{c4}{RGB}{255,127,0}

\usepackage{pgfplotstable, booktabs, colortbl}
\usetikzlibrary{arrows}

\usepackage{etoolbox}
\apptocmd{\thebibliography}{\raggedright}{}{}

\journal{Computer Physics Comunnications}

\usepackage{natbib}
\begin{document}
\begin{frontmatter}

\title{GPU parallel simulation algorithm of Brownian particles with excluded volume using Delaunay triangulations}

%% or include affiliations in footnotes:
\author[dcc]{Francisco Carter\corref{author}}
\cortext[author]{Corresponding author}
\ead{francisco.carter@ug.uchile.cl}

\author[dcc]{Nancy Hitschfeld}
\author[ua]{Crist\'obal Navarro}
\author[dfi]{Rodrigo Soto}

\address[dcc]{Department of Computer Science, FCFM, Universidad de Chile, Santiago, Chile}
\address[ua]{Institute of Informatics, Universidad Austral de Chile, Valdivia, Chile}
\address[dfi]{Physics Department, FCFM, Universidad de Chile, Santiago, Chile}

\begin{abstract}

A novel parallel simulation algorithm on the GPU, implemented in CUDA and C++, is presented for the simulation of Brownian particles that display excluded volume repulsion and interact with long and short range forces. When an explicit Euler-Maruyama integration step is performed to take into account the pairwise forces and Brownian motion, particle overlaps can appear.
The excluded volume property brings up the need for correcting these overlaps as they happen, since predicting them is not feasible due to the random displacement of Brownian particles. The proposed solution handles, at each time step, a Delaunay triangulation of the particle positions because it  allows us to efficiently solve overlaps between particles by checking just their neighborhood. The algorithm starts by generating a Delaunay triangulation of the particle initial positions on CPU, but after that the triangulation is always kept on GPU memory. We used a parallel edge-flip implementation to keep the triangulation updated during each time step, checking previously that the triangulation was not rendered invalid due to the particle displacements. We designed and implemented an exact long range force simulation  with an all-pairs $N$-body simulation, tiling the particle interaction computations based on the warp size of the target device architecture. For the short range force simulation, we developed a parallel algorithm that builds and uses Verlet lists in order to handle the particle neighborhood in parallel. The algorithm is validated with two models of active colloidal particles. Upon testing the parallel implementation of a long range forces simulation, the results show a performance improvement of up to two orders of magnitude when compared to the previously existing sequential solution. The algorithm for the short range force presents a similar performance improvement regarding the parallel long range implementation. 

\end{abstract}

\begin{keyword}
Parallel computing \sep Particle dynamics \sep Brownian dynamics \sep Overlap correction \sep Delaunay Triangulations \sep CUDA \sep GPGPU \sep N-body simulation
\end{keyword}

\end{frontmatter}

\section{Introduction}

A colloidal suspension is a mixture of microscopical  insoluble particles dispersed throughout a continuous fluid, where particle sizes range from 1~nm to 10~$\mu$m. Colloidal suspensions appear in several natural and artificial substances as the milk, mud, inks, cosmetics or latex paint, for example. Also, they are used in many intermediate industrial processes The interactions between colloidal particles of various kinds \cite{Colloids} have effects on the physical and chemical properties of the mixture such as its viscosity or light dispersion. To study these and other properties it is necessary to simulate particle systems of growing numbers ($N \geq 10^4$). Also, colloids are being used as models for active systems, to describe the motion of self-propelled microorganisms \cite{active1,active2,ABP}.

Colloids can be modelled as hard bodies subject to Brownian diffusive motion. Colloidal particles can typically  interact through the fluid in what is called hydrodynamic interactions, via electrostatic forces for charged colloids, which can be screened in an electrolyte, or with van der Waals forces \cite{Colloids}.  
In out of equilibrium conditions, phoretic forces also appear \cite{Anderson}. 
Except for the hydrodynamic forces, these interactions can be modelled with good approximation as pairwise additive forces, which in out of equilibrium conditions can eventually break the action-reaction symmetry. 

The simulation of colloidal dispersions, can be divided on two main problems executed in sequence: updating the positions of the particles due to the interparticle interactions, according to some integration rule and ensuring that the bodies do not overlap because of their movement, in order to respect the excluded volume interaction. These problems are specific instances of the n-body simulation and collision  detection respectively \cite{Navarro2014a}. In some contexts, the simulation of colloidal particles is referred as Brownian dynamics. 

There are two main methods for solving overlaps between particles: correcting all of them at once after they happen or use an event-driven approach, integrating the system until the collision instant, process the involved particles and repeat until the system reaches the target time step. The last method,  is particularly useful when inertia is important and collisions result in rebounces as in granular materials \cite{Granular,Poschel}. It requires knowing the positions of the involved bodies at the time of collision, which becomes difficult when random Brownian motion is present. For the simulation of colloidal particles, which lack of inertia and excluded volume acts  like a boundary condition rather than producing collisions, the first method is more suited. 

This work focuses on designing  and implementing a novel parallel simulation algorithm for 2D colloidal particle interacting with short and long range pairwise forces, with periodical boundaries, excluded volume and Brownian motion. The algorithm implementation takes advantage of the data-parallel computing capabilities of the GPU architecture, which have proven to be effective at accelerating the simulation process of several computational physics problems \cite{Weigel20123064, Weigel20111833, Bedorf2012}. The interactions forces are allowed to be non-reciprocal as in the case of active particles \cite{Soto2014,Soto2015}. The main contribution of this work  consists of a new and efficient method of resolving particle overlaps by using Delaunay triangulations, which are maintained periodically and fully on the graphics card. The starting positions and triangulation are initialized on the host while all the simulation code is executed on the device. The random values are also generated on the graphics card, both on the initialization and simulation phases. The algorithm uses a GPU edge-flip implementation to keep the triangulation fulfilling the Delaunay condition during each time step and to  correct inverted triangles in case they are generated  due to the particle displacements.  For the short range force simulation, we developed a parallel algorithm that builds and uses Verlet lists in order to handle the particle neighborhood in parallel. The algorithm is validated with two models of active colloidal particles. Upon testing the parallel implementation of a long range forces simulation, the results show a performance improvement of up to two orders of magnitude when compared to the previously existing sequential solution. The algorithm for the short range force presents a similar performance improvement regarding the parallel long range implementation. 

The paper is organized as follows: Section \ref{ch:preliminaries} describes the specific conditions and properties that the simulated systems must operate under. Section \ref{ch:related} lists previous related work used to solve similar problems. Section \ref{ch:overview} details the designed solution with its subcomponents, data structures, and optimizations. The implementation of the algorithm is described in Section \ref{ch:implementation}. Sections \ref{ch:results} and \ref{ch:validation} cover the tests, benchmarks, validation and used methodology, presenting the running time and performance results when compared to the other implemented solutions. Finally, section \ref{ch:discussion} rounds up the obtained results.

\section{Desctiption of the model} \label{ch:preliminaries}

This section contains the description of the involved concepts and properties of this problem that may differentiate it from other body simulation problems, such as the excluded volume and stochastic component of particle movement.

\subsection{Preliminaries}

Let $P = \{p_1, p_2, ..., p_N\}$ be a set of $N$ bodies on a \textit{d}-dimensional space. The n-body simulation is the computation of the interactions over each body in $P$, where $F_i$ corresponds to the interaction over $p_i$ by effect of $P_i = P \setminus \{p_i\}$, the set of all other bodies in the system. The interaction forces $F$ typically depend on the distance $r_{ij}$ between two bodies as $F \sim r_{ij}^{-q}$. If $q > d$ the force is said to be short ranged, while if $q \leq d$, it is considered a long range force. 
When the forces are long ranged, the set $P_i$ cannot be reduced in n-body simulations and an exact evaluation of the forces has a cost $O(N^2)$. Approximate solutions for long range interactions as the Barnes-Hut algorithm reduce the cost to $O(N \log N)$~\cite{Barnes1986}. But for short range forces, the interactions can be truncated and, therefore, $P_i$ can be reduced to the neighborhood of particles close to $p_i$. In this case, the evaluation of the forces costs $O(N * N_{NL})$ on average, where $N_{NL}$ is the average number of neighbors of a body \cite{Yao2004}. Since the construction of the list is $O(N^2)$ for evaluating all pairwise distances between bodies, it is possible to partition the simulation domain in cells so that close bodies get binned together in the same cells. Assignment of bodies to their respective cells takes $O(N)$ time~\cite{Allen,Frenkel}.

The simulation domain is a two-dimensional $L \times L$, periodical box across the $X$ and $Y$ axes, meaning that the particles wrap around the box as they move across its boundaries. For the distance calculations  between particles, including force calculations, we follow the minimum-image convention, in which a particle interacts with another via its real position or its image depending on which is the shortest.

\subsection{Particle interaction without excluded volume}
Microscopic particles move in an overdamped regime, with no inertia. When subject to a force $\vec{F}$, the equation of motion is simplified to $d\vec{r}/dt= \gamma \vec{F}$, where $\gamma$ is the mobility. Absorbing the mobility coefficient into the force, which will then  have velocity units, in a time step $\Delta t$, the integration rule for updating the position of a particle over time is performed using the Euler-Maruyama method:
\begin{equation} \label{eq:integration}
\vec{r}_i(t + \Delta t) = \vec{r}_i(t) + \vec{F}_i(t) \Delta t + \vec{\xi} \sqrt{D \Delta t},
\end{equation}
where $\vec{F}_i$ is the deterministic velocity obtained from the interactions between the particle $i$ and $P_i$, $D$ is the diffusion coefficient, and $\vec{\xi}$ is a random vector, where the components follow a normal distribution of zero mean and unit variance, and corresponds to a noise added that takes into account the diffusive Brownian motion \cite{Miguel2000}.

The force model we use for the simulations describes the interaction of self-diffusiophoretic active particles \cite{Soto2014}. In this model, particles can be of different type, characterized by two charges, $\alpha$ and $\mu$; the former is responsible of creating the concentration field, while the second describes the response of a particle to the field, leading to the following interaction law:
\begin{equation}
\vec{F}_i = \sum_{k \neq i} \mu_i \alpha_k \vec{f}(\vec{r}_i - \vec{r}_k),
\end{equation}
where $\vec{f}(\vec{r}) = {\vec{r}}/{r^3}$ for the studied long range force, while $\vec{f}(\vec{r}) = {\vec{r}}/{r^7}$ for the short range interaction. %Both long range and short range forces can coexist in the same simulation. 
Note that if $\alpha_i\neq\mu_i$, the  action-reaction symmetry is broken and self-motion is possible. Charged colloidal particles are included in this model if $\alpha_i=\mu_i=q_i$, equal to the electric charge of the particles.

\begin{figure}[H] \centering
	\includegraphics{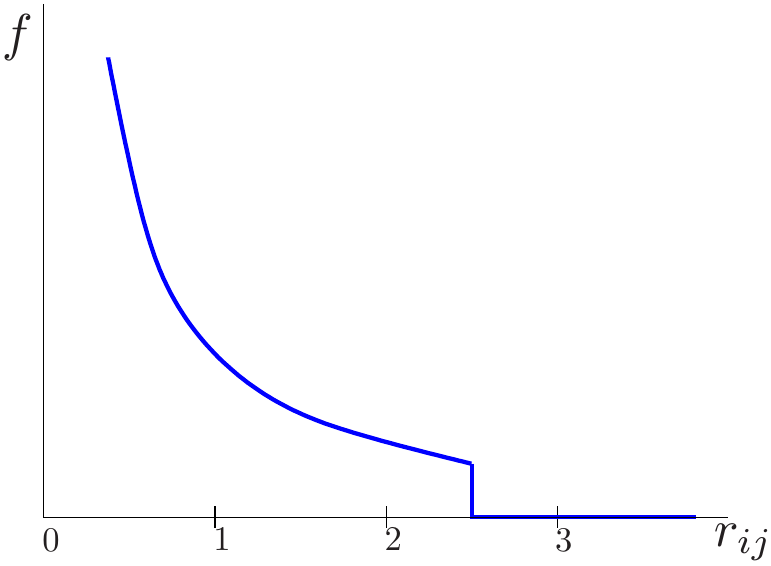}
	\caption{ Cutoff of the short range forces. For distances larger than $r_\text{cutoff}$, the force is small and therefore is set to zero to speed up calculations. The jump at  $r_\text{cutoff}$ has been exaggerated for illustration purposes. }
	\label{fig:cutoff}
\end{figure}

Since the short range force decays much faster with distance compared to the long range force, its calculation considers a cutoff radius from which the value of the force is considered zero, as shown on Figure \ref{fig:cutoff}. The  short range force is then computed as:
\begin{equation} \label{eq:displacement}
	\vec{F}_{ij}(\vec{r}_{ij})= 
	\begin{cases}
		\mu_i \alpha_k \vec{f}(\vec{r}_{ij}),& \text{if } r_{ij}\leq r_\text{cutoff}\\
	    0, & \text{otherwise}
	\end{cases}
\end{equation}
We used $r_\text{cutoff} = 2.5\sigma$ for the simulated short range force in our experiments, where $\sigma$ is the particle diameter.

\subsection{Excluded volume}

The simulated particles are represented as hard disks with a uniform diameter $\sigma$. Although here we consider only monodisperse colloids, it is direct to extend the method to polydisperse systems  where radii dot not differ too much. Since the integration rule (\ref{eq:integration}) ignores the excluded volume condition, it can happen that the updated positions produce overlaps between two or more particles, resulting in a physical impossibility. In order to ensure this property, at the end of each time step, the members of all  overlapping pairs $(p_i, p_j)$ are moved apart from each other in a way that corrects the overlaps:
\begin{align} \label{eq:overlaps}
	\vec{r}_1\prime &= \vec{r}_1 - \delta^* \frac{ (\vec{r}_2 - \vec{r}_1) } { |\vec{r}_{12}| } & \vec{r_2}\prime &= \vec{r}_2 - \delta^* \frac{ (\vec{r}_1 - \vec{r}_2) } { |\vec{r}_{12}| },
\end{align}
where $\vec{r}_1$ and $\vec{r}_2$ are the original positions and $\vec{r}_1\prime, \vec{r}_2\prime$  the updated positions. If $\delta^* = (\sigma - |\vec{r_{12}}|)/2$, the particles would move in opposite directions from each other along $\hat{n} = (\vec{r}_2 - \vec{r}_1) / |\vec{r}_{12}|$, leaving the particles in tangential contact. If $\delta^* = \sigma - |\vec{r}_{12}|$, the movement is proportional to the magnitude of the previously existing overlap, simulating a bounce effect resulting from the collision at some instant $t^* \leq t + \Delta t$. This last value is the one used for processing the overlaps in the simulation and guarantees that no accumulation is produced at the contact distance.
 
\subsection{Stochastic displacements}

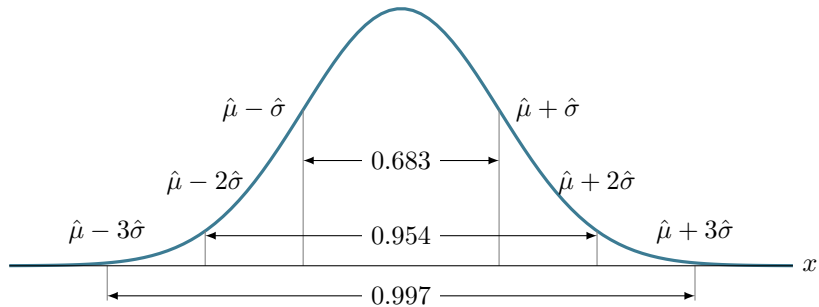
\begin{figure}[H] \centering
	\begin{tikzpicture}
	\begin{axis}[
		no markers, 
		domain=0:8, 
		samples=100,
		ymin=0,
		axis lines*=left, 
		xlabel=$x$,
		every axis y label/.style={at=(current axis.above origin),anchor=south},
		every axis x label/.style={at=(current axis.right of origin),anchor=west},
		height=5cm, 
		width=12cm,
		xtick=\empty, 
		ytick=\empty,
		enlargelimits=false, 
		clip=false, 
		axis on top,
		grid = major,
		hide y axis]
		
		\addplot [very thick,cyan!50!black] {gauss(x, 4, 1)};
		\pgfmathsetmacro\valueA{gauss(2,4,1)}
		\pgfmathsetmacro\valueB{gauss(3,4,1)}
		\draw [gray] (axis cs:2,0) -- (axis cs:2,\valueA)
		(axis cs:6,0) -- (axis cs:6,\valueA);
		\draw [gray] (axis cs:3,0) -- (axis cs:3,\valueB)
		(axis cs:5,0) -- (axis cs:5,\valueB);
		\draw [yshift=1.4cm, latex-latex](axis cs:3, 0) -- node [fill=white] {$0.683$} (axis cs:5, 0);
		\draw [yshift=0.4cm, latex-latex](axis cs:2, 0) -- node [fill=white] {$0.954$} (axis cs:6, 0);
		
		\pgfmathsetmacro\valueC{gauss(1,4,1)}
		\pgfmathsetmacro\valueD{gauss(2.5,4,1)}
		
		\draw [gray] (axis cs:1,0) -- (axis cs:1,-\valueA)
		(axis cs:7,0) -- (axis cs:7,-\valueA);
		\draw [yshift=-0.4cm, latex-latex](axis cs:1, 0) -- node [fill=white] {$0.997$} (axis cs:7, 0);
		\node at (axis cs:1, \valueA)  {$\hat\mu - 3\hat\sigma$}; 
		\node at (axis cs:7, \valueA)  {$\hat\mu + 3\hat\sigma$};
		\node at (axis cs:2, \valueD)  {$\hat\mu - 2\hat\sigma$}; 
		\node at (axis cs:6, \valueD)  {$\hat\mu + 2\hat\sigma$};
		\node at (axis cs:2.5, \valueB)  {$\hat\mu - \hat\sigma$}; 
		\node at (axis cs:5.5, \valueB)  {$\hat\mu + \hat\sigma$};						
	\end{axis}
\end{tikzpicture}
	\caption{Normal distribution with mean $\hat\mu$ and standard deviation $\hat\sigma$. The probability to get numbers in the ranges $[\hat\mu-\hat\sigma,\hat\mu+\hat\sigma]$, $[\hat\mu-2\hat\sigma,\hat\mu+2\hat\sigma]$, and $[\hat\mu-3\hat\sigma,\hat\mu+3\hat\sigma]$ are $0.683$, $0.954$, and $0.997$, respectively.}
	\label{fig:normal}
\end{figure}

\begin{figure}[H]
	%\subcaptionbox{Rerolling values out of range.}
	{\begin{tikzpicture}
	\begin{axis}[
		no markers, 
		domain=0:6, 
		samples=100,
		ymin=0,
		axis lines*=left, 
		every axis y label/.style={at=(current axis.above origin),anchor=south},
		every axis x label/.style={at=(current axis.right of origin),anchor=west},
		height=5cm,
		width=12cm,
		xtick=\empty,
		ytick=\empty,
		enlargelimits=false, 
		clip=false, 
		axis on top,
		grid = major,
		hide y axis,
		xmin = -1, xmax = 7]
		
		\addplot [name path=f,very thick,magenta!50!black]
		{gauss(x, 3, 1) + 0.02};
		\path[name path=axis] (axis cs:0,0) -- (axis cs:6,0);
		\pgfmathsetmacro\valueA{gauss(1,4,1)  + 0.02}
		\draw [very thick,magenta!50!black]
		(axis cs:0,0) -- (axis cs:0,\valueA)
		(axis cs:6,0) -- (axis cs:6,\valueA);
		\addplot[thick,color=red,fill=red, fill opacity=0.05]
		fill between[of=f and axis];
		\node[below] at (axis cs:0, 0) {\scalebox{2}{$\hat\mu - 3\hat\sigma$}}; 
		\node[below] at (axis cs:6, 0) {\scalebox{2}{$\hat\mu + 3\hat\sigma$}};	
	\end{axis}
\end{tikzpicture}}
	\hfill
	%\subcaptionbox{Truncating the generated values.}
	{\begin{tikzpicture}
	\begin{axis}[
		no markers, 
		domain=0:6, 
		samples=100,
		ymin=0,
		axis lines*=left, 
		every axis y label/.style={at=(current axis.above origin),anchor=south},
		every axis x label/.style={at=(current axis.right of origin),anchor=west},
		height=5cm,
		width=12cm,
		xtick=\empty, 
		ytick=\empty,
		enlargelimits=false, 
		clip=false, 
		axis on top,
		grid = major,
		hide y axis,
		xmin = -1, xmax = 7]
		
		\addplot [name path=f,very thick,cyan!50!black]
		{gauss(x, 3, 1)};
		\path[name path=axis] (axis cs:1,0) -- (axis cs:7,0);
		\pgfmathsetmacro\valueA{gauss(2,4,1)}
		\pgfmathsetmacro\valueB{gauss(1,4,1)}
		\draw [very thick,cyan!50!black]
		(axis cs:0,0) -- (axis cs:0,\valueA)
		(axis cs:6,0) -- (axis cs:6,\valueA);
		\addplot[thick,color=blue,fill=blue, fill opacity=0.05]
		fill between[of=f and axis];
		\node[below] at (axis cs:0, 0) {\scalebox{2}{$\mu - 3\sigma$}}; 
		\node[below] at (axis cs:6, 0) {\scalebox{2}{$\mu + 3\sigma$}}; 
	\end{axis}
\end{tikzpicture}}
	\caption{Distributions that result after discarding values larger than $3\hat\sigma$ from the original Gaussian distribution. Two methods are used. Left: Rerolling values out of range. Right: Truncating the generated values. In this case, Dirac-delta contributions of small amplitude appear at $\hat\mu\pm3\hat\sigma$. }
	\label{fig:gaussianoptions}
\end{figure}
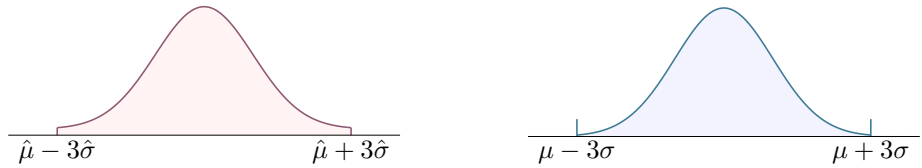

The particle displacements on (\ref{eq:integration}) have a random noise component $\vec{\xi}$, modeled as a random variable with standard normal  (or Gaussian) distribution with zero mean and standard deviation $\hat\sigma=\sqrt{D\Delta t}$. 
Reducing $\Delta t$, the deterministic and stochastic displacements in each time step are reduced. However, for a Gaussian distribution, it is always possible that large values are generated (at the tail of the distribution), leading to excessively large displacements (see Figure \ref{fig:normal}). To avoid these problems, the simulation ignores values larger than 3 standard deviations. We considered two methods in order to achieve this, as shown on Figure \ref{fig:gaussianoptions}:
\begin{enumerate}
	\item[(a)] Reroll the values outside the range $[- 3 \hat\sigma,  3 \hat\sigma]$.
	\item[(b)] Truncate the values to the range $[ - 3 \hat\sigma,  3 \hat\sigma]$.
\end{enumerate} 

Both methods produce probability distributions different from each other and from the original; while the first alternative raises the probability of all values in range, the second one raises the probability at the edges. These modifications do not generate a noticeable  statistic distortion, since the considered range includes $99.7\%$ of the possible values. In our simulations, we opted for the second method, which turns out to be faster and better suited for parallel execution, since it needs to generate a single random number instead of a variable quantity of random values in the first method.

\section{Related work} \label{ch:related}

For short-range forces calculation, the standard technique is the use of Verlet lists \cite{Verlet1967,Allen,Frenkel}. The authors in \cite{Lipscomb2012, Proctor2013} parallelize the list construction by having a $O(N^2)$ list of all possible pairs of bodies. A predicate checking closeness between the pair members is evaluated over all elements of the list, which can then be used for a key-value sort to group all the neighboring pairs consecutively in the array. A parallel scan operation allows to get the number of elements that must be copied to the neighbor list. The authors then combine this algorithm with fixed cell partitioning in order to replace distance calculations with less-expensive cell neighborhood checks.

For the parallel n-body simulation, with full calculation of the $O(N^2)$ forces, Nyland et al.~\cite{Nyland2007} developed a grid-style tiling algorithm, reading the particles from the global space and storing them on GPU shared memory, increasing performance as multiple threads read from that space at a lower latency. Partitioning the load/store process on groups of $p$ particles allows fitting an arbitrary input size on the hardware-limited shared memory size. Burtscher and Pingali~\cite{Burtscher2011} parallelized the Barnes-Hut simulation~\cite{Barnes1986}, which computes an approximation for the force, representing the cell hierarchy k$d$-tree as multiple arrays for each node field. It uses atomic lock operations to build the tree in parallel, throttling the threads that failed to get the lock so they do not waste bandwidth with unsuccessful lock requests. The tree is then filled with the center of mass data, starting from lower nodes in the tree according to the order of allocation for the scan. Bedorf et al.~\cite{Bedorf2012} uses a Z-order curve to sort the particles spatially. Each thread is assigned to a particle, applying a mask value to it to determine the octree cell the particle should be assigned. The linking of the tree is made by assigning a thread to each cell node and then doing a binary search over the corresponding Z-order key to find both the parent and child nodes, if appropiate.

To detect and process collisions, Hawick and Playne~\cite{Hawick2012} developed a multi-GPU algorithm with a tiling scheme similar to the one used by Nyland et al.~\cite{Nyland2007}. If a pair of particles overlap, the associated threads store the index of its colliding neighbor and the time at which the collision occurred. The collisions then are resolved iteratively starting from the earliest, redoing the previous process in order to find possible new collisions. 

Finally, for overlap correction, Strating~\cite{Strating1999} describes a brute-force sequential algorithm that checks all pairs of bodies for possible overlaps and corrects them following equation (\ref{eq:overlaps}). The algorithm may need to iterate an unbounded number of times at each time step because some corrections may generate new overlaps with neighboring particles. 

\section{Algorithm} \label{ch:overview}

This section describes the parallel algorithm in detail. It includes the generation of the initial data, data structures,  overlapping detection and   correction, and Delaunay condition updates, among others, putting emphasis in what threads are doing at each time step.

\subsection{Overview}

The simulation consists of two phases: (i) sequential initialization of the simulation data, followed by a host to device transfer and (ii) a parallel simulation phase. The initial positions are initialized over a triangular mesh with $N^*\geq N$ vertex,  where each vertex represents a particle and their types are assigned randomly according to the specified concentrations. A sample of $N$ particles is selected from the mesh by Reservoir Sampling~\cite{Vitter1985}, resulting in the input particle set, which is homogeneous in space. 

The n-body algorithm for the long range force is based on a grid-style tiling, which uses the shared memory of the multiprocessor assigned to each thread block to store the particles in groups. In this algorithm each thread is mapped to exactly one particle in the system, and since it is possible to lack action-reaction symmetry on the force, no redundant computation is done unlike the cases where $\vec{f}_{ij} = -\vec{f}_{ji}$. 

Once the forces are calculated, the particles are advanced one time step using eqn (\ref{eq:integration}). As a result, particle overlaps can appear. When $\Delta t$ is small enough, for hard disks of similar or equal radii, only neighbor particles can overlap. Then, to detect and correct overlaps, instead of a brute-force algorithm that would check all $O(N^2)$ pairs, only neighbors are checked. The Delaunay triangulation~\cite{DeBerg2008} is particularly well suited to detect neighbors for monodisperse or slightly polydisperse disks. In dense systems, the overlap corrections can be highly non-local, as the correction of on pair can generate a sequence of other overlaps that need correction. It is therefore not clear a priori the computational cost of this stage, which is the reason why we consider both short and long range interaction forces.

The Delaunay triangulation can be built constructively or from an existing triangulation. Lawson's algorithm \cite{Lawson1972} accepts a triangulation as input and transforms it into a Delaunay triangulation via a finite sequence of edge-flip operations \cite{DeBerg2008}. Based on the Lawson algorithm, Navarro et al.~\cite{Navarro2014} developed a parallel implementation for generating quasi-Delaunay triangulations, so it is possible to keep the triangulation updated without need of host-device memory transfers. These are quasi-Delaunay because exact predicates are too expensive on the GPU; nevertheless, this approximate  construction is sufficient for our problem. 
However, since the input for the Navarro et al.~algorithm  must be a valid triangulation, we must first correct potential triangle inversions with invalid edge intersections, which can result from the particle displacements. With this strategy, the Delaunay triangulation of the particle positions is built only once from scratch on the host \cite{cgal:eb-16a,cgal:k-pt2-13-16a}, which is transferred to the device. Thereafter, the triangulation is maintained updated after each time step on the device.
%Finally, note that upon considering the Periodic Boundary Conditions, the resulting triangulation is representable as a flat torus surface \cite{cgal:k-pt2-13-16a} with both major and minor radii equal to $L$.

\begin{algorithm}[H]
	\caption{Particle system simulation}
	\begin{algorithmic}[1]
		\Require $P = \{p_1, ..., p_N\}$ list of particle positions
		\Ensure $P = \{p_1, ..., p_N\}$ list of positions updated to current time
		\Procedure {runSimulation}{$P$}
		\State Generate starting position of $N$ particles
		\State Build the Delaunay triangulation 
		\For{$t \gets 0$ \textbf{to} $T_f$}
		\State Integrate the $N$ particles on $t + \Delta t$
		\State Correct inverted triangles
		\State Update Delaunay triangulation
		\State Correct overlaps between particles
		\EndFor
		\EndProcedure
	\end{algorithmic}
\end{algorithm}

\subsection{Data structures}

We store the simulation data as a Structure of Arrays (SoA) on global device memory, using total $O(N)$ space. The particle data consists of their position $(x_i, y_i)$ and their charges $(\alpha_i, \mu_i)$, stored as floating point vector types \footnote{\texttt{float2} or \texttt{double2}.} in order to increase bandwidth utilization \cite{NVIDIA2015}. We use an additional buffer array for positions so that writes are not done at the same adresses for reads, avoiding a synchronization step. We store the simulation parameters that remain unchanged during a same instance on a constant device memory structure \cite{NVIDIA2015}, such as $N, D, \Delta t, \sigma$ and derived constants $\sigma^2$ and $\sqrt{D \Delta t}$. Additionally, we store the triangulation data using the same scheme as \cite{Navarro2014}.

\subsection{Inverted triangle detection}

There are two possible reasons for a triangle inversion: an edge gets inverted because the distance vector between its terminal vertex changes orientation or because one of its vertex crosses the edge opposite to it. The first problem means that the particles went through each other, which is a physical impossibility and must not be allowed, and can only takes place for large $\Delta t$. The criterion  $\vec{r}_{ij}^0 \cdot \vec{r}_{ij}^1 < 0$ is used to determine if the above situation happened on the current time step, where $\vec r_{ij}=\vec r_j-\vec r_i$ is the distance vector between the compared particles, and $\vec{r}_{ij}^0$ and $\vec{r}_{ij}^1$ are evaluated with current and previous positions, respectively. The buffer array allows to compare the distances before and after integration, and the edges of the triangulation show the pairs that need checking. Once an invalid movement is detected, the last positions are discarded and integration is repeated with a lower $\Delta t$ value than the currently used. 

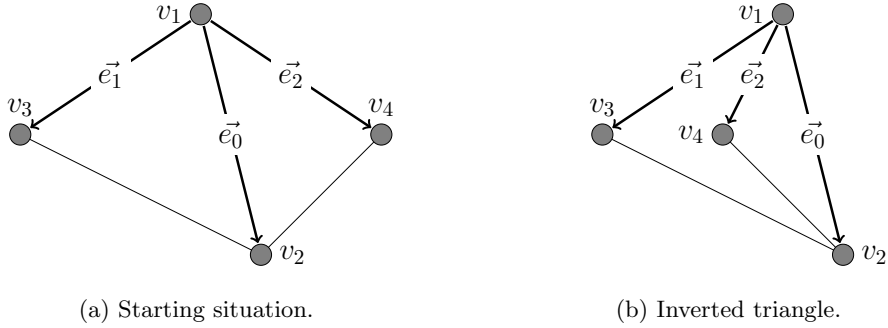
\begin{figure}[H] \centering
	\subcaptionbox{Starting situation.}
	{
		\begin{tikzpicture}
    \tikzstyle{every node}=[font=\Large]

	\coordinate (op1) at (0, 0);
	\coordinate (op2) at (6, 0);
	\coordinate (e1) at (3, 2);
	\coordinate (e2) at (4, -2);
		
	\draw {
		(op1) node[circle, draw, fill=black!50, inner sep=0pt, minimum width=10pt, label=\Large $v_3$]  {}
		(op2) node[circle, draw, fill=black!50, inner sep=0pt, minimum width=10pt, label=\Large $v_4$] {}
		(e1) node[circle, draw, fill=black!50, inner sep=0pt, minimum width=10pt, label=left:\Large $v_1$] {}
		(e2) node[circle, draw, fill=black!50, inner sep=0pt, minimum width=10pt, label=right:\Large $v_2$] {}
		(op1) -- (e2) -- (op2)
	};
		
	\draw[->,very thick, shorten >=0.2cm, shorten <=0.2cm] (e1) -- (e2) node[midway, fill=white] {\Large $\vec{e_0}$};
	\draw[->,very thick, shorten >=0.2cm, shorten <=0.2cm] (e1) -- (op1) node[midway, fill=white] {\Large $\vec{e_1}$};
	\draw[->,very thick, shorten >=0.2cm, shorten <=0.2cm] (e1) -- (op2) node[midway, fill=white] {\Large $\vec{e_2}$};
\end{tikzpicture}
	}
	\hfill
	\subcaptionbox{Inverted triangle. \label{fig:inverted_triangle}}
	{
		\begin{tikzpicture}
	\coordinate (op1) at (0, 0);
	\coordinate (op2) at (2, 0);
	\coordinate (e1) at (3, 2);
	\coordinate (e2) at (4, -2);
	
	\draw {
		(op1) node[circle, draw, fill=black!50, inner sep=0pt, minimum width=10pt, label=\Large $v_3$] {}
		(op2) node[circle, draw, fill=black!50, inner sep=0pt, minimum width=10pt, label=left:\Large $v_4$] {}
		(e1) node[circle, draw, fill=black!50, inner sep=0pt, minimum width=10pt, label=left:\Large $v_1$] {}
		(e2) node[circle, draw, fill=black!50, inner sep=0pt, minimum width=10pt, label=right:\Large $v_2$] {}
		(op1) -- (e2) -- (op2)
	};
	
	\draw[->,very thick, shorten >=0.2cm, shorten <=0.2cm] (e1) -- (e2) node[midway, fill=white] {\Large $\vec{e_0}$};
	\draw[->,very thick, shorten >=0.2cm, shorten <=0.2cm] (e1) -- (op1) node[midway, fill=white] {\Large $\vec{e_1}$};
	\draw[->,very thick, shorten >=0.2cm, shorten <=0.2cm] (e1) -- (op2) node[midway, fill=white] {\Large $\vec{e_2}$};
\end{tikzpicture}	
	}
	\caption{Inverted triangle detection using barycentric coordinates. Particle sizes are scaled down compared to distances.}
	\label{fig:barycentric}
\end{figure}

Finally, the inverted triangle detection becomes equal to checking if a vertex crossed (towards a neighboring triangle) any of the edges that enclose it, which is equivalent to point-in-triangle detection. 
Using the barycentric coordinates $d$, $s$, and $t$ on triangles (see Figure \ref{fig:barycentric}), 
\begin{align} 
    \vec{e}_0 &= \vec{v}_2 - \vec{v}_1 & d_2 &= \vec{e}_2 \times \vec{e}_0 \\
    \vec{e}_1 &= \vec{v}_3 - \vec{v}_1 & s_2 &= \vec{e}_1 \times \vec{e}_0 \\
    \vec{e}_2 &= \vec{v}_4 - \vec{v}_1 & t_2 &= \vec{e}_2 \times \vec{e}_1
\end{align}

the criterion used to detect the edges that must be flipped is
\begin{equation} \label{eq:predicate}
	\text{flip}(\mathbf{v_1}, \mathbf{v_2}, \mathbf{v_3}, \mathbf{v_4}) = 
	\begin{cases}
	    (s_2 \leq 0) \land (t_2 \leq 0) \land (s_2 + t_2 \geq d_2),& \text{if } d_2 < 0\\
	    (s_2 \geq 0) \land (t_2 \geq 0) \land (s_2 + t_2 \leq d_2),& \text{otherwise}
    \end{cases}
\end{equation}

Figure \ref{fig:barycentric} shows the vectors used on the predicate that checks if the original edge must be flipped. The predicate becomes true when applied to edge $(v_1, v_2)$ in (\subref{fig:inverted_triangle}), because point $v_4$ lies inside the triangle $(v_1, v_2, v_3)$. This means that $(v_1, v_2)$ must be replaced with $(v3_1, v_4)$, as in a common edge-flip operation. In this case, $v_3$ and $v_4$ are stored as opposite vertices to edge $(v_1, v_2)$ in the triangulation data structure in GPU device memory.

\begin{figure}[H] \centering
	\subcaptionbox{Starting situation.}
	{
		\begin{tikzpicture}[thick]
    \tikzstyle{every node}=[circle, draw, fill=black!50, inner sep=0pt, minimum width=10pt, font=\Huge]

	\coordinate (a) at (0, 0);
	\coordinate (b) at (6, 0);
	\coordinate (c) at (3, 2);
	\coordinate (d) at (4, -2);
	\coordinate (e) at (8, 0.5);
	
	\draw {
		(a) node[label=$a$] {}
		(b) node[label=$b$] {}
		(c) node[label=left:$c$] {}
		(d) node[label=right:$d$] {}
		(e) node[label=$e$] {}
		(a) -- (d) -- (c) -- cycle
		(b) -- (c) -- (d) -- cycle
		(b) -- (c) -- (e) -- cycle
		(b) -- (d) -- (e) -- cycle
	};
\end{tikzpicture}
	}
	\hfill
	\subcaptionbox{Particle $b$ moves over the edge \textbf{bc}.}
	{
			
\begin{tikzpicture}[thick]
    \tikzstyle{every node}=[circle, draw, fill=black!50, inner sep=0pt, minimum width=10pt, font=\Huge]

	\coordinate (a) at (0, 0);
	\coordinate (b) at (2, 0);
	\coordinate (c) at (3, 2);
	\coordinate (d) at (4, -2);
	\coordinate (e) at (8, 0.5);
	
	\draw[fill=cyan] (b) -- (c) -- (d) -- cycle;
	\draw {
		(a) node[label=$a$] {}
		(b) node[label=$b$] {}
		(c) node[label=left:$c$] {}
		(d) node[label=right:$d$] {}
		(e) node[label=$e$] {}
		(a) -- (d) -- (c) -- cycle
		(b) -- (c) -- (d) -- cycle
		(b) -- (c) -- (e) -- cycle
		(b) -- (d) -- (e) -- cycle
	};
	
\end{tikzpicture}
	}

	\subcaptionbox{Edge flip between \textbf{ab} and \textbf{cd}.}
	{
		\begin{tikzpicture}[thick]
    \tikzstyle{every node}=[circle, draw, fill=black!50, inner sep=0pt, minimum width=10pt, font=\Huge]

    \tkzDefPoint(0, 0){A}
    \tkzDefPoint(2, 0){B}
    \tkzDefPoint(3, 2){C}
    \tkzDefPoint(4, -2){D}
    \tkzDefPoint(8, 0.5){E}
    \tkzDefPoint(0, -5){F}
    
    % circumcircle
    \tkzCircumCenter(B,C,E)\tkzGetPoint{G}
    %\tkzDrawPoint(G)
    \tkzDrawCircle[color=cyan](G,B)
    
    \tkzCircumCenter(B,D,E)\tkzGetPoint{F}
    \tkzDrawCircle[color=magenta](F,B)

	\coordinate (a) at (0, 0);
	\coordinate (b) at (2, 0);
	\coordinate (c) at (3, 2);
	\coordinate (d) at (4, -2);
	\coordinate (e) at (8, 0.5);
	
	\draw {
		(a) node[label=$a$] {}
		(b) node[label=$b$] {}
		(c) node[label=left:$c$] {}
		(d) node[label=right:$d$] {}
		(e) node[label=$e$] {}
		(a) -- (d) -- (b) -- cycle
		(a) -- (b) -- (c) -- cycle
		(b) -- (c) -- (e) -- cycle
		(b) -- (d) -- (e) -- cycle
	};
	
\end{tikzpicture}
	}
	\hfill
	\subcaptionbox{Edge flip between \textbf{cd} and \textbf{be}.}
	{
		\begin{tikzpicture}[thick]
    \tikzstyle{every node}=[circle, draw, fill=black!50, inner sep=0pt, minimum width=10pt, font=\Huge]
	
    \tkzDefPoint(0, 0){A}
    \tkzDefPoint(2, 0){B}
    \tkzDefPoint(3, 2){C}
    \tkzDefPoint(4, -2){D}
    \tkzDefPoint(8, 0.5){E}
    \tkzDefPoint(0, -5){Z}
    
    % circumcircle
    \tkzCircumCenter(B,C,D)\tkzGetPoint{G}
    %\tkzDrawPoint(G)
    \tkzDrawCircle[color=cyan](G,C)
    
    \tkzCircumCenter(C,D,E)\tkzGetPoint{F}
    \tkzDrawCircle[color=magenta](F,C)
    
	\coordinate (a) at (0, 0);
	\coordinate (b) at (2, 0);
	\coordinate (c) at (3, 2);
	\coordinate (d) at (4, -2);
	\coordinate (e) at (8, 0.5);
	
	\draw {
		(a) node[label=$a$] {}
		(b) node[label=$b$] {}
		(c) node[label=left:$c$] {}
		(d) node[label=right:$d$] {}
		(e) node[label=$e$] {}
		(a) -- (d) -- (b) -- cycle
		(a) -- (b) -- (c) -- cycle
		(b) -- (c) -- (d) -- cycle
		(c) -- (d) -- (e) -- cycle
	};
	
\end{tikzpicture}
	}
	\caption{Inverted triangle correction. The cyan shaded triangle was inverted by the movement of particle $b$. Particle sizes are scaled down compared to distances.}
	\label{fig:inverted_tri1}
\end{figure}

It is worth noting on Figure \ref{fig:inverted_tri1} that the movement of point $b$ across edge $(c,d)$ creates an intersection between it and edge $(b,e)$. The edge flip between $(a,b)$ and $(c,d)$ removes the inverted triangle, restoring the local triangulation. The triangulation may still not satisfy the Delaunay property, so additional edge flips may be needed on further steps. For this stage, we use the Navarro et al. algorithm \cite{Navarro2014}.

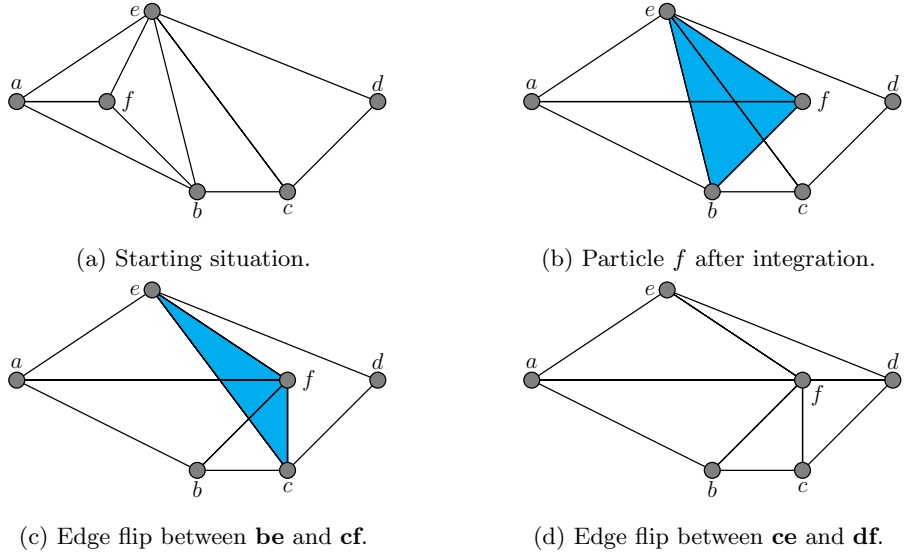
\begin{figure}[H] \centering
	\subcaptionbox{Starting situation.}
	{
		\begin{tikzpicture}[thick]
    \tikzstyle{every node}=[circle, draw, fill=black!50, inner sep=0pt, minimum width=10pt, font=\Large]

	\coordinate (a) at (0, 0);
	\coordinate (b) at (4, -2);
	\coordinate (c) at (6, -2);
	\coordinate (d) at (8, 0);
	\coordinate (e) at (3, 2);
    \coordinate (f) at (2, 0);
	
	\draw {
		(a) node[label=$a$] {}
		(b) node[label=below:$b$] {}
		(c) node[label=below:$c$] {}
		(d) node[label=$d$] {}
		(e) node[label=left:$e$] {}
		(f) node[label=right:$f$] {}
		(a) -- (b) -- (f) -- cycle
		(a) -- (e) -- (f) -- cycle
		(b) -- (c) -- (e) -- cycle
		(c) -- (d) -- (e) -- cycle
	};
\end{tikzpicture}
	}
	\hfill
	\subcaptionbox{Particle $f$ after integration.}
	{
		\begin{tikzpicture}[thick]
    \tikzstyle{every node}=[circle, draw, fill=black!50, inner sep=0pt, minimum width=10pt, font=\Large]

	\coordinate (a) at (0, 0);
	\coordinate (b) at (4, -2);
	\coordinate (c) at (6, -2);
	\coordinate (d) at (8, 0);
	\coordinate (e) at (3, 2);
    \coordinate (f) at (6, 0);
	
	\draw[fill=cyan] (b) -- (e) -- (f) -- cycle;
	
	\draw {
		(a) node[label=$a$] {}
		(b) node[label=below:$b$] {}
		(c) node[label=below:$c$] {}
		(d) node[label=$d$] {}
		(e) node[label=left:$e$] {}
		(f) node[label=right:$f$] {}
		(a) -- (b) -- (f) -- cycle
		(a) -- (e) -- (f) -- cycle
		(b) -- (c) -- (e) -- cycle
		(c) -- (d) -- (e) -- cycle
	};
\end{tikzpicture}
	}
    \vspace{0.5cm}
	\subcaptionbox{Edge flip between \textbf{be} and \textbf{cf}.}
	{
		\begin{tikzpicture}[thick]
    \tikzstyle{every node}=[circle, draw, fill=black!50, inner sep=0pt, minimum width=10pt, font=\Large]

	\coordinate (a) at (0, 0);
	\coordinate (b) at (4, -2);
	\coordinate (c) at (6, -2);
	\coordinate (d) at (8, 0);
	\coordinate (e) at (3, 2);
    \coordinate (f) at (6, 0);
	
	\draw[fill=cyan] (c) -- (e) -- (f) -- cycle;
	
	\draw {
		(a) node[label=$a$] {}
		(b) node[label=below:$b$] {}
		(c) node[label=below:$c$] {}
		(d) node[label=$d$] {}
		(e) node[label=left:$e$] {}
		(f) node[label=right:$f$] {}
		(a) -- (b) -- (f) -- cycle
		(a) -- (f) -- (e) -- cycle
		(b) -- (c) -- (f) -- cycle
		(c) -- (e) -- (f) -- cycle
		(c) -- (d) -- (e) -- cycle
	};
\end{tikzpicture}
	}
	\hfill
	\subcaptionbox{Edge flip between \textbf{ce} and \textbf{df}.}
	{
		\begin{tikzpicture}[thick]
    \tikzstyle{every node}=[circle, draw, fill=black!50, inner sep=0pt, minimum width=10pt, font=\Large]

	\coordinate (a) at (0, 0);
	\coordinate (b) at (4, -2);
	\coordinate (c) at (6, -2);
	\coordinate (d) at (8, 0);
	\coordinate (e) at (3, 2);
    \coordinate (f) at (6, 0);
	
	\draw {
		(a) node[label=$a$] {}
		(b) node[label=below:$b$] {}
		(c) node[label=below:$c$] {}
		(d) node[label=$d$] {}
		(e) node[label=left:$e$] {}
		(f) node[label=below right:$f$] {}
		(a) -- (b) -- (f) -- cycle
		(a) -- (f) -- (e) -- cycle
		(b) -- (c) -- (f) -- cycle
		(c) -- (d) -- (f) -- cycle
		(d) -- (e) -- (f) -- cycle
	};
\end{tikzpicture}
	}
	\caption{Inverted triangle correction with two edge flips. Particle sizes are scaled down compared to distances.}
	\label{fig:inverted_tri2}
\end{figure}

Figure \ref{fig:inverted_tri2} shows a situation where particle $f$ moves across edges $(b,e)$ and $(c,e)$, needing two consecutive flips in order to restore the local triangulation. While function \ref{eq:predicate} cannot properly evaluate this case or similar movements across further distances, this kind of inversion can be detected allowing to revert the step. Anyway, small time steps guarantee that this situation is extremely unlikely to happen.

\subsection{Overlap correction}

The overlap correction uses the topological information contained in the edges of the Delaunay triangulation, which allows for each particle fast access to the neighborhood of particles that may be overlapping with it. The algorithm maps threads to edges in such a way that each thread handles one edge of the triangulation. A thread gets the positions of the particles that form the edge, checking if there exists an overlap between them ($r_{ij}<\sigma$). If the check is positive, the algorithm computes the displacements of the involved particles according to (\ref{eq:displacement})
Since the same particle can be part of many edges at once, the algorithm sums atomically the displacements in a global array, in order to avoid concurrency hazards.

\begin{algorithm}[H]
	\caption{Overlap correction}
	\begin{algorithmic}[1]
		\Require $P_0$ starting positions, $E$ triangulation edges
		\Ensure $P_1$ displacements over each particle
		\Procedure {correctOverlaps}{$P_0, P_1, E$}
		\For{$i \gets 0$ \textbf{to} $|E|$}
		\State $e_i \gets E[i]$ 
		\State $b_i \gets P_0[e_i.\texttt{first}]$
		\State $b_j \gets P_0[e_i.\texttt{second}]$
		\State $r_{ij} \gets \texttt{dist}(b_i, b_j)$
		\If{$r_{ij} < \sigma$}
		\State $\delta \gets \sigma - r_{ij}$ 
		\State \texttt{atomicAdd}($P_1[e_i.x]$, $-\delta\, \vec r_{ij}$)
		\State \texttt{atomicAdd}($P_1[e_i.y]$, $\delta\,  \vec r_{ij}$)
		\EndIf
		\EndFor
		\EndProcedure
	\end{algorithmic}
\end{algorithm}

Once the algorithm computes the total displacements, it maps each thread with a particle in the same way as described previously for edges. Each thread then updates the position of its particle, applying the periodic boundary conditions when necessary. It is possible that the updated positions may still have some of the previous overlaps or even have some newly generated ones. In this case,  the algorithm repeats the previous process until no overlaps are present.

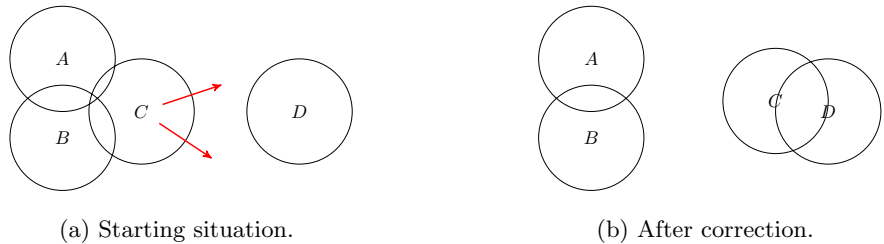
\begin{figure}[H] \centering
	\subcaptionbox{Starting situation.}
	{
		\begin{tikzpicture}
	\coordinate (a) at (0, 2.5);
	\coordinate (b) at (0, 1);
	\coordinate (c) at (1.5, 1.5);
	\coordinate (d) at (4.5, 1.5);
		
	\draw
	{
	    (a) circle (1cm) node {$A$}
	    (b) circle (1cm) node {$B$}
	    (c) circle (1cm) node {$C$}
	    (d) circle (1cm) node {$D$}
	};
	\draw [->, red, thick, >=stealth', shorten >=-1.6cm, shorten <=2.2cm] (a) -- (c);
	\draw [->, red, thick, >=stealth', shorten >=-1.6cm, shorten <=2cm] (b) -- (c);
\end{tikzpicture}
		
	}
	\hspace{2cm}
	\subcaptionbox{After correction.}
	{
		\begin{tikzpicture}
	\coordinate (a) at (0, 2.5);
	\coordinate (b) at (0, 1);
	\coordinate (c) at (3.5, 1.7);
	\coordinate (d) at (4.5, 1.5);
		
	\draw 
	{
	    (a) circle (1cm) node {$A$}
	    (b) circle (1cm) node {$B$}
	    (c) circle (1cm) node {$C$}
	    (d) circle (1cm) node {$D$}
	};
\end{tikzpicture}
		
	}
	\caption{Possible instability with the parallel overlap correction. The displacement over $C$ caused by the overlaps with $A$ and $B$ can be greater than needed,	which can generate a larger overlap with a neighboring particle $D$. This problem is prevented by truncating the displacements to a maximum amount.}
	\label{fig:overlaps}
\end{figure}

When adding the partial displacements on a particle, it may happen, if these point in the same direction, that the resulting total displacement is excessively large (see Figure \ref{fig:overlaps}). These displacements, larger to what is needed to solve the overlap, can generate new overlaps. Eventually, the correction of the new overlaps can result in an instability, where the displacements increase with alternating sign and the iterative procedure does not converge. A solution is truncating the displacement with the heuristic value $\sigma / 4$ (half the particle radius), preventing the emergence of the instability.
 
Finally, it is worth mentioning that the parallel correction algorithm presented here does not correspond to  a parallelization of the sequential algorithm of Strating \cite{Strating1999}, which displaces particles sequentially, while in our case the displacements are added and performed in parallel. Hence, due to the chaotic dynamics of the system, the small differences in these algorithms will produce different outputs for finite $\Delta t$.

\subsection{Long range forces} \label{ch:improvements}

An improvement to the long range force calculation consists on using the intrinsic warp shuffle instruction, which allows a thread access to the registers of other threads belonging to the same warp. Each thread is assigned a lane number that identifies it from the other warp members, allowing them to read different particles from global memory. Then, each warp member takes turns in propagating the data of its corresponding particle to the other threads, who can read it via the \texttt{\_shfl()} instruction by passing as argument the lane number of the thread currently in turn. Once the whole warp has shared the data among its members, each member reads a particle from global memory and repeats the same process until all particles have been visited. The main advantage of this optimization is a greater efficiency of memory accesses, since most of the time the threads are sharing data at registry level instead of more expensive load requests on global memory. Also, the concurrent execution of the warp members makes unnecessary the explicit synchronization of the inner warp shuffle loop.

\begin{algorithm}[H]
	\caption{Particle system integration}
	\begin{algorithmic}[1]
		\Require $P_0$ particle array $(x_i, y_i, \alpha_i, \mu_i)$
		\Ensure $P_1$ particles with updated positions
		\Procedure {integrate}{$P_0, P_1$}
		\For{$i \gets 0$ \textbf{to} $|P|$}
    		\State $l_i \gets \texttt{threadIdx.x}$ \& $(\texttt{warpSize}-1)$
    		\State $\vec{b}_i \gets P_0[i]$ 
    		\State $\vec{v}_i \gets 0$
    		\For{$j \gets 0$ \textbf{to} $|P|$; $j \gets j+\texttt{warpSize}$}
        		\State $\vec{b}_j \gets P_i[j + l_i]$
        		\For{$k \gets 0$ \textbf{to} \texttt{warpSize}}
            		\State $\vec{b}_k \gets \texttt{\_shfl}(\vec{b}_j, k)$
            		\State $\vec{r}_{ik} \gets \texttt{dist}(\vec{b}_i, \vec{b}_k)$
            		\State $\vec{v}_i \gets \vec{v}_i - \vec{r}_{ik} \cdot (\alpha_i \mu_k)/r_{ik}^3$
        		\EndFor
        		\State \texttt{\_syncthreads()}
    		\EndFor
		\EndFor
		\State $P_1[i].x \gets \vec{b}_i.x + \vec{v}_i.x \Delta t + \vec{\xi}_i.x \sqrt{D \Delta t}$ 
		\State $P_1[i].y \gets \vec{b}_i.y + \vec{v}_i.y \Delta t + \vec{\xi}_i.y \sqrt{D \Delta t}$ 
		\EndProcedure
	\end{algorithmic}
\end{algorithm}

The $.x$ and $.y$ operators reference the data of the vectorized CUDA structures for each respective variable. For example, the noise $\vec{\xi_i}$ has $\hat{x}$ and $\hat{y}$ components, so it is grouped as a single vector for increased memory performance \cite{NVIDIA2015}.

\section{Implementation} \label{ch:implementation}

The  parallel algorithms described in the previous section were implemented on CUDA 7.5 and C++ 11, using function templates to choose between \texttt{float} and \texttt{double} precision formats at compile time. We use the CGAL library \cite{cgal:eb-16a} to create the 2D periodic Delaunay triangulation, which is then sent to device memory alongside the particle data before starting the simulation. The random numbers used on the parallel implementations when initializing the starting positions and generating noise during integration are created with the XORWOW pseudorandom number generation algorithm of the \texttt{cuRAND} library \cite{iNVI15a}, using the host and device APIs respectively. Each configuration has two particle types, although the program can support a variable number of particle types for simulating. For comparison purposes, we also implemented a fully sequential long range forces algorithm with the overlap correction discussed on \cite{Strating1999}, and a parallel short range forces algorithm using Verlet lists and a discrete grid over the simulation box. The neighbor list computation during the Verlet lists construction in parallel is similar to the one described on \cite{green2010}, grouping together all the particles that belong in the same cells.

When the simulation finishes, the final positions are brought back to host memory and written to an output file. The visualizations on Figures \ref{fig:configurations}, \ref{fig:locality}, \ref{fig:abp1} and \ref{fig:abp2} were generated reading the respective output files. 

\section{Performance results} \label{ch:results}

\paragraph{Parameters:} We generated inputs for 11 different values of $N$ and 5 parameter configurations, as described on Table \ref{tbl:config}. The starting positions are generated semi-randomly as described on section \ref{ch:overview}, keeping the same seed value for the random number generator across all simulation instances.

\begin{table}[H] \centering

\pgfplotstabletypeset[
	col sep=comma,
	every even row/.style={ before row={\rowcolor[gray]{0.9}} },
	every head row/.style={ before row=\toprule,after row=\midrule },
	every last row/.style={ after row=\bottomrule},
	columns/conc1/.style={column name=$\phi_1$},
	columns/alpha1/.style={column name=$\alpha_1$},
	columns/mu1/.style={column name=$\mu_1$},
	columns/conc2/.style={column name=$\phi_2$},
	columns/alpha2/.style={column name=$\alpha_2$},
	columns/mu2/.style={column name=$\mu_2$},
	columns/d/.style={column name=$\rho$}
] {data/config_values}

	\caption{Parameters used for the tests, where each configuration is identified by a digit and all of them contain two types of particles. $\phi_i$ is the concentration of particles of type $i$, $\alpha_i, \mu_i$ are the charges used in the force calculation, and $\rho$ is the packing fraction of particles on the simulation box.}
	\label{tbl:config}
\end{table}

Each configuration has two types of particles with charges $\alpha_i, \mu_i$. The fraction of particles of each type is given by $\phi_i=N_i/N$, where $N_i$ is the number of particles of each type and $N=N_1+N_2$ is the total number of particles. The area fraction $\rho=N (\sigma/2)^2/L^2$ is a measure of the particle density. To study scaling times, we change the length of the simulation box to keep density constant when increasing $N$:
\begin{align}
	L(N) &= \sqrt{\frac{N \pi (\sigma / 2)^2}{\rho}}
\end{align}
The values for $N$ start at $2^{10}$, raising the exponent by 1 until $N = 2^{20}$. Finally, we kept constant the values for $\sigma=1, \Delta t = 0.01, D = 0.01, \delta = 1.0$ for all configurations and input sizes. 

\begin{figure}
	\subcaptionbox{Configuration 0}{\includegraphics[scale=0.15]{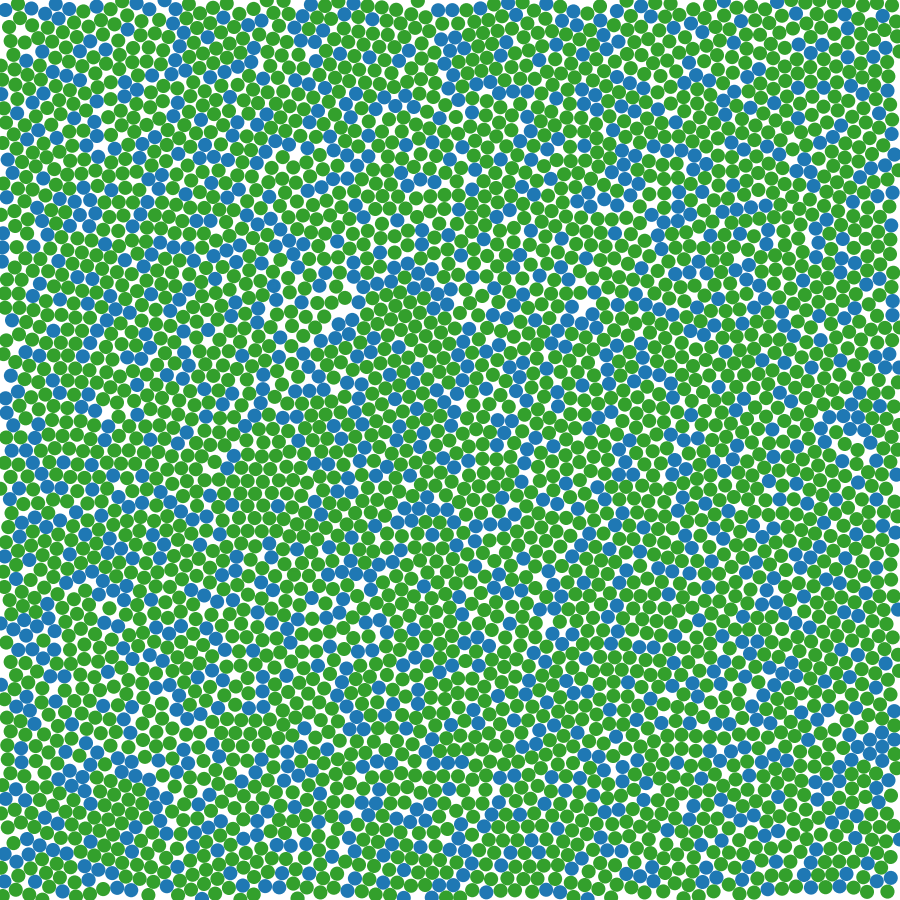}}
	\hfill
	\subcaptionbox{Configuration 1}{\includegraphics[scale=0.15]{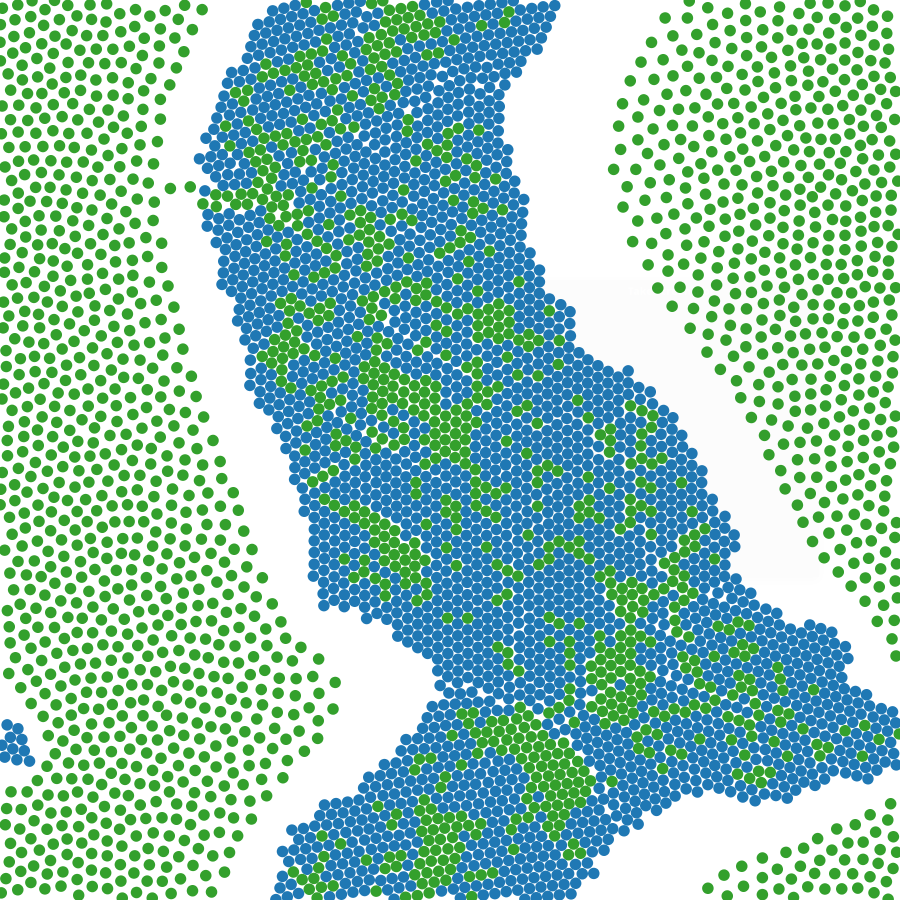}}
    
	\subcaptionbox{Configuration 2}{\includegraphics[scale=0.15]{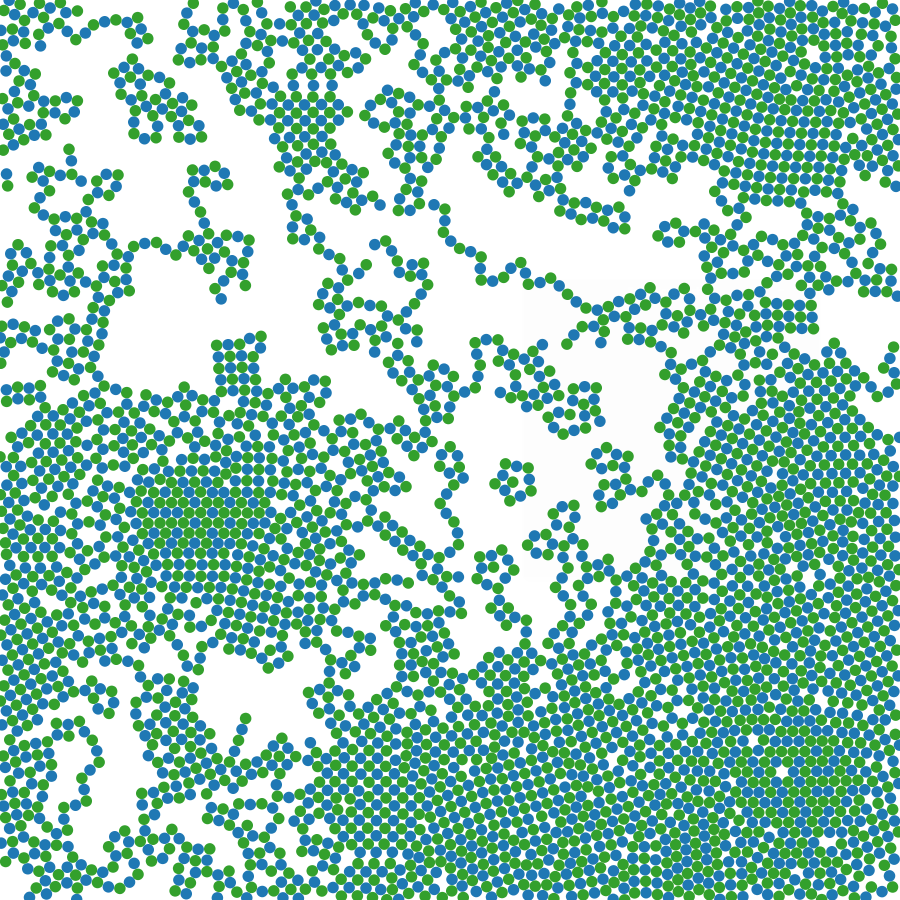}}
	\hfill
    \subcaptionbox{Configuration 3}{\includegraphics[scale=0.15]{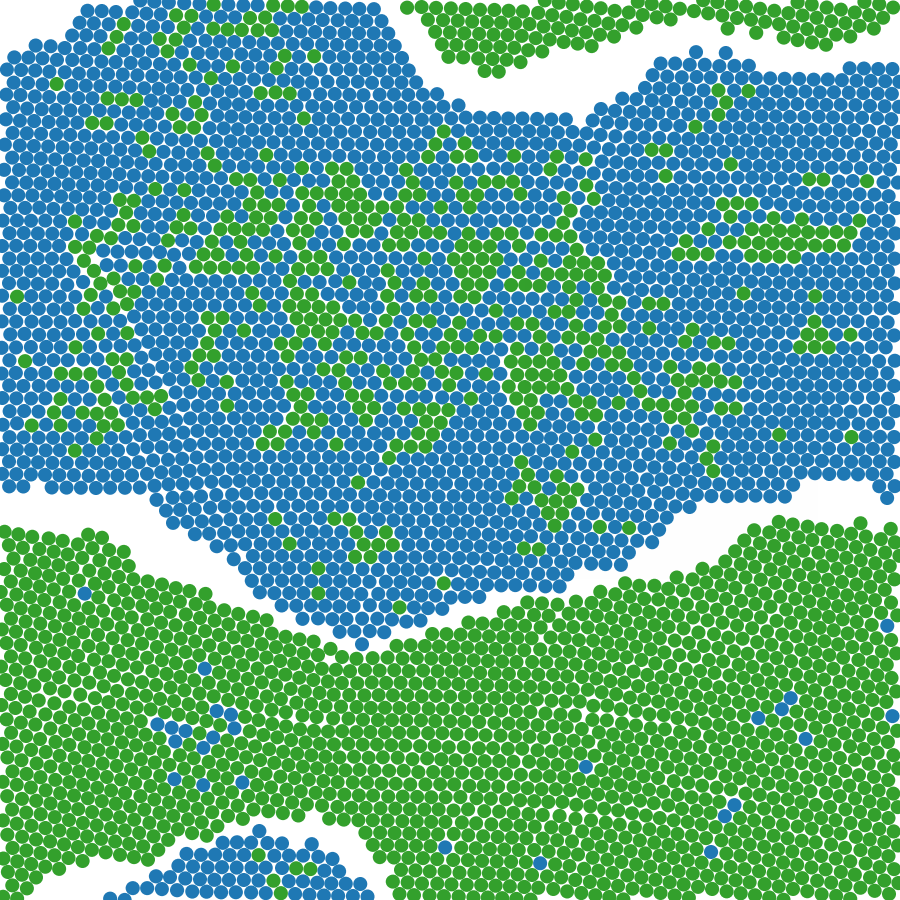}}
    
    \centering
    \subcaptionbox{Configuration 4}{\includegraphics[scale=0.15]{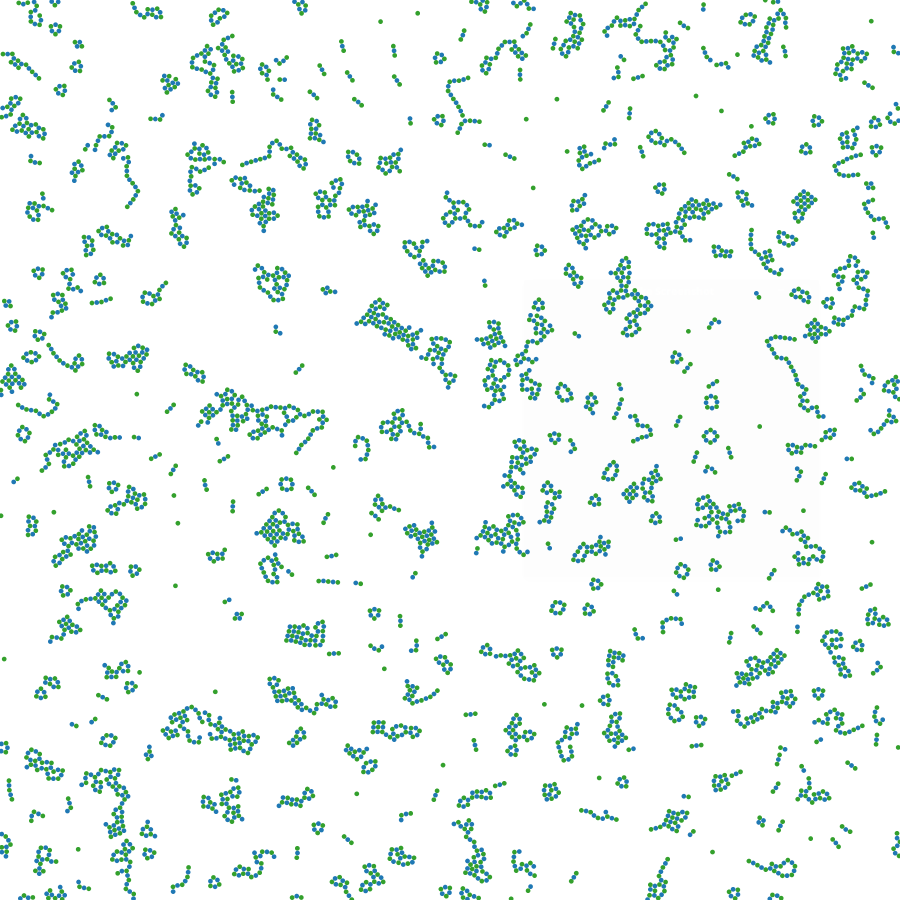}}
	\caption{Particle positions after $10^4$ iterations, with $N = 4096$ and $\Delta t = 0.01$. Type 1 particles are in green and type 2 in blue. The configurations and particle types for each system are described on Table \ref{tbl:config}. }
	\label{fig:configurations}
\end{figure}

Figure \ref{fig:configurations} displays the particle positions after $10^4$ time steps for each configuration. The election of the parameters  help to test the algorithms under different conditions of fluidity, density and homogeneity. For c0, there is an asymmetric attraction between particles of type 1 and 2, resulting in an homogeneous mixture, with fluid-like motion. In c1, there is a larger concentration of type 2 particles, which self-attract forming a dense cluster, segregated from type 1 particles, which self-repel forming in a gas-like state. In configuration c2, equal particles repel, while dissimilar particles attract, favouring the formation of chain-like structures, where 1 and 2 particles alternate. In c3, the situation is the opposite, where equal particles attract, while dissimilar particles repel, leading to the formation of dense segregated clusters. Finally, the interactions in c4 are analogous to those of c2, in a dilute regime, resulting in the formation of small clusters. 

\paragraph{System:} We ran the tests on a machine with a Tesla K40c GPU and a Intel(R) Xeon(R) CPU E5-2640 v3 @ 2.60GHz. The tests for both the sequential and parallel implementations were made on the same machine.

\paragraph{Compilation:} We compiled the program using \texttt{nvcc V7.5.17} with compiler options \texttt{--std=c++11 -gencode arch=compute\_30,code=sm\_30}. For the sequential code, we used \texttt{g++ 4.9.2} with options \texttt{--std=c++11 -O3}.

\paragraph{Metric:} 
We ran simulations for 100 iterations, long enough to ensure that particle collisions happen frequently, except for the first iterations where the bodies are separated from each other. There we compute the average execution time and iteration averages for overlap correction cycles and edge flips.

\begin{figure}[H] \centering
	\subcaptionbox{Long range forces.}{\begin{tikzpicture}
	\begin{loglogaxis}[
	legend pos=north west,
	xlabel= \# particles,
	ylabel= Time (ms) ]
		\addplot[color=c0,mark=star] table[x=n, y=c0, col sep=comma] {data/delaunay_time};
		\addplot[color=c1,mark=*] table [x=n, y=c1, col sep=comma] {data/delaunay_time};
		\addplot[color=c2,mark=square*] table [x=n, y=c2, col sep=comma] {data/delaunay_time};
		\addplot[color=c3,mark=triangle*] table [x=n, y=c3, col sep=comma] {data/delaunay_time};
		\addplot[color=c4,mark=diamond*] table [x=n, y=c4, col sep=comma] {data/delaunay_time};
		\legend{c0, c1, c2, c3, c4}
	\end{loglogaxis}
\end{tikzpicture}
		}
	\hfill
	\subcaptionbox{Short range forces.}{\begin{tikzpicture}
	\begin{loglogaxis}[
	legend pos=north west,
	legend columns=2, 
xlabel= \# particles,
	ylabel= Time (ms) ]
		\addplot[color=c0,mark=star] table[x=n, y=c0, col sep=comma] {data/verlet_time};
		\addplot[color=c1,mark=*] table [x=n, y=c1, col sep=comma] {data/verlet_time};
		\addplot[color=c2,mark=square*] table [x=n, y=c2, col sep=comma] {data/verlet_time};
		\addplot[color=c3,mark=triangle*] table [x=n, y=c3, col sep=comma] {data/verlet_time};
		\addplot[color=c4,mark=diamond*] table [x=n, y=c4, col sep=comma] {data/verlet_time};	 
		\legend{c0, c1, c2, c3, c4}
	\end{loglogaxis}
\end{tikzpicture}
		}
	\caption{Average execution time per time step for simulations using long and short range forces, using the parameters shown on Table \ref{tbl:config}.}
	\label{plot:execution_time}
\end{figure}
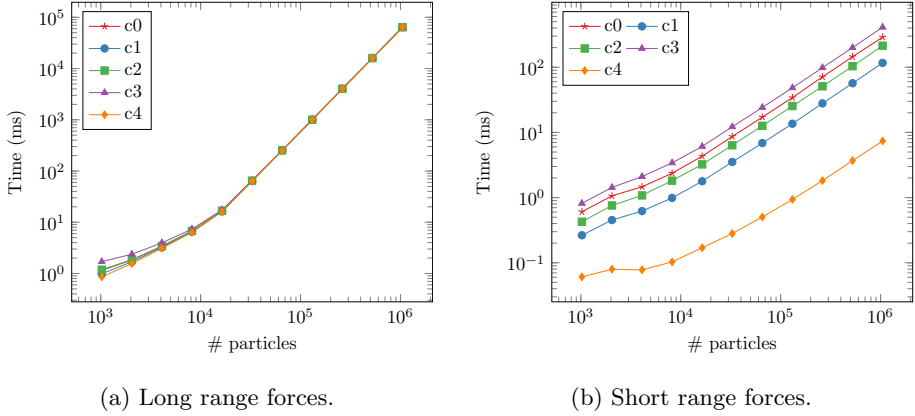

The average execution time per time step, presented in Figure \ref{plot:execution_time}, shows two interesting features. First, for the case with long range forces the execution time is $O(N^2)$, while for short range forces it is $O(N)$. Since both include the overlap correction algorithm, this implies that the execution time for the later is $O(N)$. Second, except for small systems, in the case of long range forces the execution time does not depend on the configuration, while for the short range forces, there is a clear dependence, with increasing complexity for $c_4, c_1, c_2, c_0, c_3$ (the same order of complexity is observed for long range forces at small $N$). This result is consistent with a cost $O(N)$ for the overlap correction, with a prefactor that may depend on the density and extension of the clusters. 

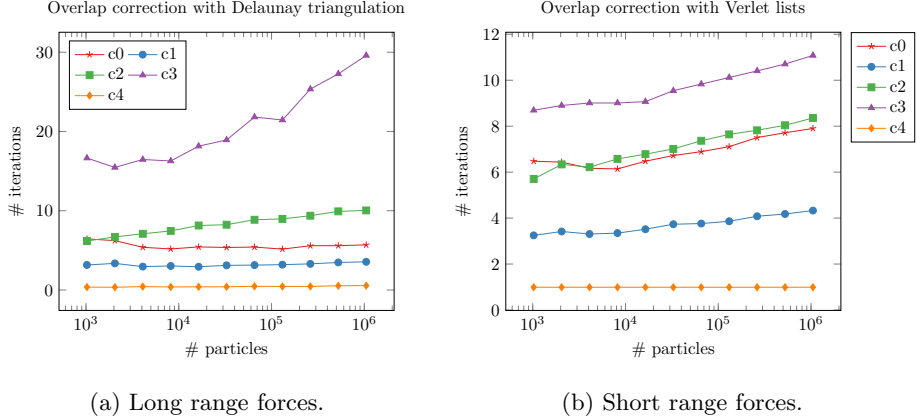
\begin{figure}[H] \centering
	\subcaptionbox{Long range forces.}{\begin{tikzpicture}
	\begin{semilogxaxis}[title= Overlap correction with Delaunay triangulation,
	legend pos=north west,
	legend columns=2, 
	xlabel= \# particles,
	ylabel= \# iterations ]
		\addplot[color=c0,mark=star] table[x=n, y=c0, col sep=comma] {data/delaunay_overlaps};
		\addplot[color=c1,mark=*] table [x=n, y=c1, col sep=comma] {data/delaunay_overlaps};
		\addplot[color=c2,mark=square*] table [x=n, y=c2, col sep=comma] {data/delaunay_overlaps};
		\addplot[color=c3,mark=triangle*] table [x=n, y=c3, col sep=comma] {data/delaunay_overlaps};
		\addplot[color=c4,mark=diamond*] table [x=n, y=c4, col sep=comma] {data/delaunay_overlaps};
		\legend{c0, c1, c2, c3, c4}
	\end{semilogxaxis}
\end{tikzpicture}
		}
	\hfill
	\subcaptionbox{Short range forces.}{\begin{tikzpicture}
	\begin{semilogxaxis}[title= Overlap correction with Verlet lists,
	legend pos=outer north east, 
	xlabel= \# particles,
	ylabel= \# iterations ]
		\addplot[color=c0,mark=star] table[x=n, y=c0, col sep=comma] {data/verlet_overlaps};
		\addplot[color=c1,mark=*] table [x=n, y=c1, col sep=comma] {data/verlet_overlaps};
		\addplot[color=c2,mark=square*] table [x=n, y=c2, col sep=comma] {data/verlet_overlaps};
		\addplot[color=c3,mark=triangle*] table [x=n, y=c3, col sep=comma] {data/verlet_overlaps};
		\addplot[color=c4,mark=diamond*] table [x=n, y=c4, col sep=comma] {data/verlet_overlaps};
		\legend{c0, c1, c2, c3, c4}
	\end{semilogxaxis}
\end{tikzpicture}
		}
	\caption{Average overlap correction iterations per time step for short and long range forces simulation, using the parameters shown on Table \ref{tbl:config}.}
	\label{plot:overlaps}
\end{figure}

To study the dependence of different configurations on the complexity of the overlap correction, in Figure \ref{plot:overlaps} we plot the average per time step of iterations needed to correct all overlaps. The increasing complexity for $c_4, c_1, c_2, c_0, c_3$ is consistent with the previous results on Figure \ref{plot:execution_time}, because in $c_3$ most of the particles participate in corrections, while in $c_2$ almost half of the particles are excluded due to repulsion between same-type particles. However, this does not explain why $c_4$ has the least complexity factor, even though half of the particles overlap. This happens because all the overlaps on $c_4$ are corrected on the first iteration, which is probably due to the small size of the clusters. The number of iterations follow the same order in complexity as the execution time. Except for $c_4$ where clusters are disconnected, the number of iterations grow with $N$. This effect is due to the percolation of the large clusters, which cover the entire box and, therefore, the corrections become non-local and system size dependent. This growth is nevertheless weak, following an approximate logarithmic law. It is also noteworthy that the curves for long range forces are constant on $c_4, c_1$ and $c_0$, slightly grow on $c_2$ and is relatively greater on $c_3$.

\begin{figure}[H] \centering
	\begin{tikzpicture}
	\begin{semilogxaxis}[
	title= Edge-flip iterations per time step,
	legend pos=north west,
	legend columns=2, 
	xlabel= \# particles,
	ylabel= \# iterations ]
		\addplot[color=c0,mark=star] table[x=n, y=c0, col sep=comma] {data/delaunay_flips};
		\addplot[color=c1,mark=*] table [x=n, y=c1, col sep=comma] {data/delaunay_flips};
		\addplot[color=c2,mark=square*] table [x=n, y=c2, col sep=comma] {data/delaunay_flips};
		\addplot[color=c3,mark=triangle*] table [x=n, y=c3, col sep=comma] {data/delaunay_flips};
		\addplot[color=c4,mark=diamond*] table [x=n, y=c4, col sep=comma] {data/delaunay_flips};
		\legend{c0, c1, c2, c3, c4}
	\end{semilogxaxis}
\end{tikzpicture}
		
	\caption{Average edge-flip iterations per time step for long range forces integration, using the simulation parameters shown on Table \ref{tbl:config}. The short range forces algorithm is not analyzed, since it does not use Delaunay triangulations.}
	\label{fig:edge_flips}
\end{figure}
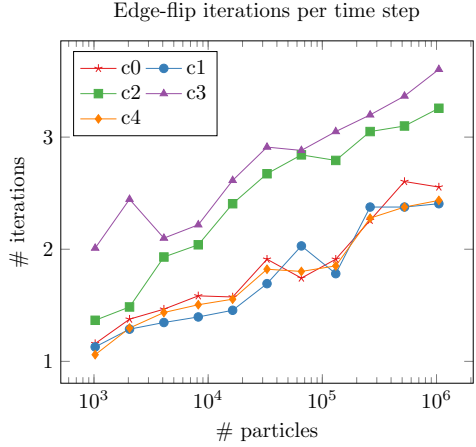

We also measured the performance of the Delaunay triangulation update algorithm, reporting the average of edge flip iterations made for both inverted triangle corrections and Lawson's algorithm. The curves obtained on Figure \ref{fig:edge_flips} are less regular than the previous results, but keep the same general tendency. Unlike the curves for overlap correction, on where $c_4$ shows much smaller values than the other configurations, here the curve is comparable to $c_0$ and $c_1$. This happens because the underlying triangulation for $c_4$ has a great number of slivers, formed by the small particle density that forms relatively long edges. Then, according to the inverted triangle condition on section \ref{ch:overview}, it is more likely for $c_4$ to produce inverted triangle than the other configuration, whose triangles are more equilateral. The average for edge-flip iterations for long range forces has linear growth for all configurations, noting that $c_2$ and $c_3$ are the hardest cases to solve, as is the case on Figure \ref{plot:overlaps}. Though the number of iterations grows with $N$, it still remains negligible regarding the total time of a time step, so it is not a priority target for optimization.

Finally, we compared the n-body algorithms for long range forces, used on the different implementations without considering overlap corrections. \texttt{shuffle} is the presented optimization using warp-shuffle, while \texttt{sharedMem} is the GPU device memory algorithm described on section \ref{ch:overview}, observing a performance improvement of up to $2.4$ times from optimizing the parallel implementation for all tested values of $N$. It is also noteworthy that the optimized n-body
algorithm allows simulation of $N = 10^6$ particles at the same time that the sequential implementation solves the problem for $10^5$ bodies. For input sizes relevant to this study ($N \geq 10^4$), the time used by the sequential implementation is two orders of magnitude higher than the parallel solution, which allows the simulation of bigger particle systems for a longer physical time.

\begin{figure}[H] \centering
	\begin{tikzpicture}
	\begin{loglogaxis}[title= Integration of a time step,
	legend pos=north west, 
	xlabel= \# particles,
	ylabel= Time (ms) ]
		\addplot[color=c0,mark=star]
		table[x=n, y=shfl, col sep=comma] {data/nbody_comp};
		\addplot[color=c1,mark=*]
		table [x=n, y=shr, col sep=comma] {data/nbody_comp};
		\addplot[color=c2,mark=square*]
		table [x=n, y=cpu, col sep=comma] {data/nbody_comp};
		\legend{shuffle, sharedMem, CPU}
	\end{loglogaxis}
\end{tikzpicture}
		
	\caption{Comparison between execution times in milliseconds for both implementations of the quadratic n-body algorithm, using configuration $c_0$ described on Table \ref{tbl:config}.}
\end{figure}
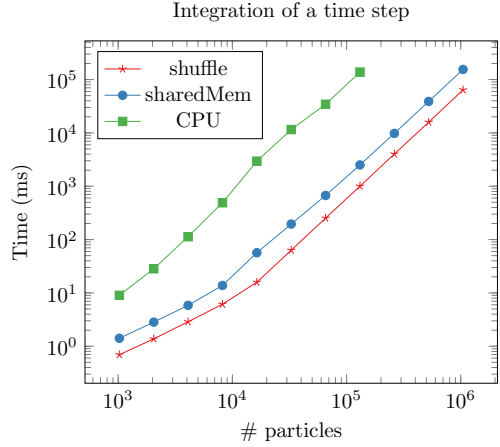

\section{Validation} \label{ch:validation}

\begin{figure}[H]
	\subcaptionbox{$\Delta t = 10^{-2}$}{\includegraphics[scale=0.12]{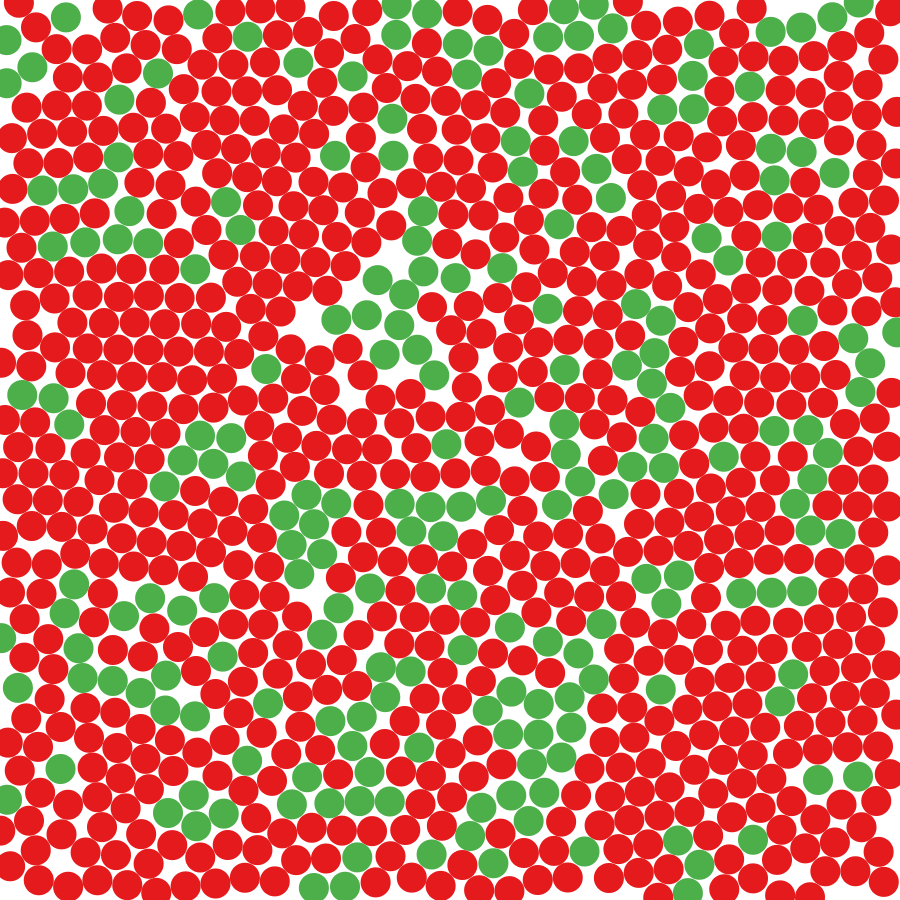}}
	\hfill
	\subcaptionbox{$\Delta t = 10^{-3}$}{\includegraphics[scale=0.12]{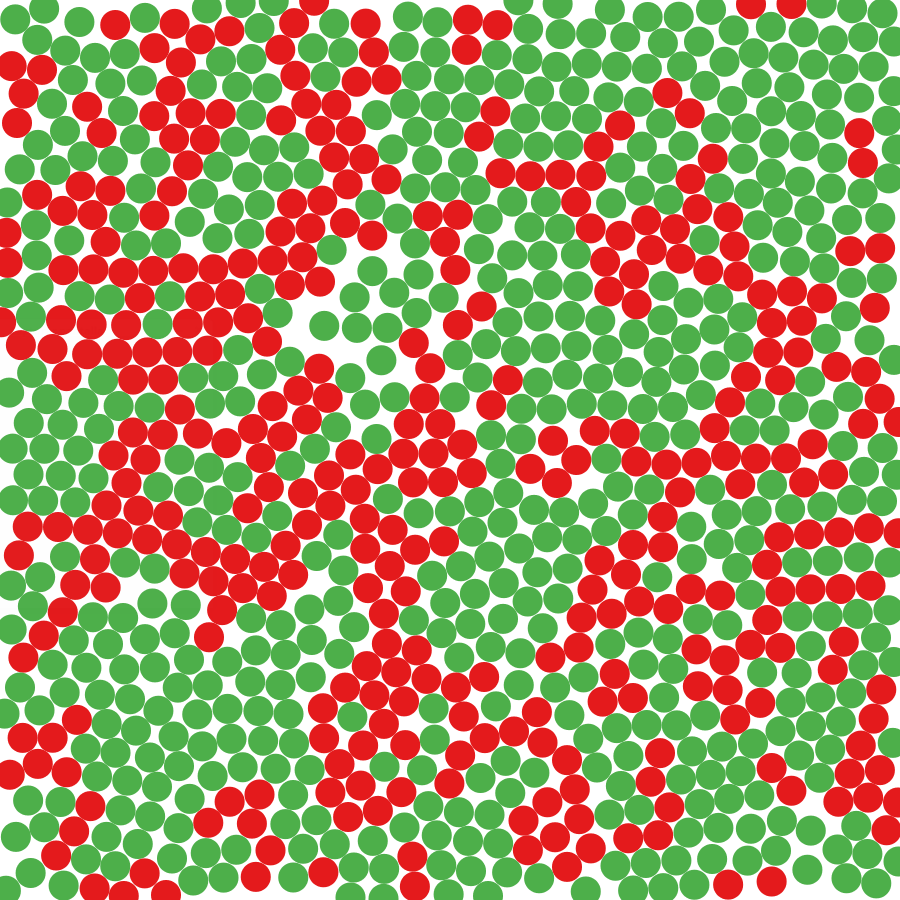}}
	\hfill
	\subcaptionbox{$\Delta t = 10^{-4}$}{\includegraphics[scale=0.12]{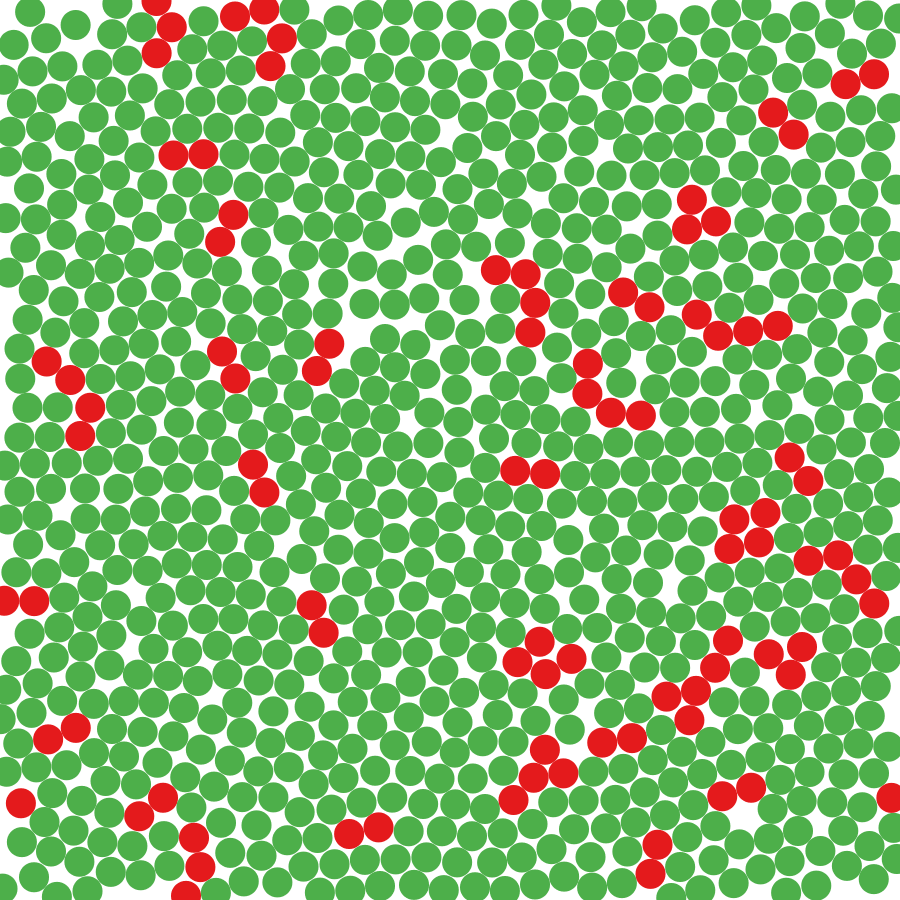}}
	
	\subcaptionbox{$\Delta t = 10^{-2}$}{\includegraphics[scale=0.12]{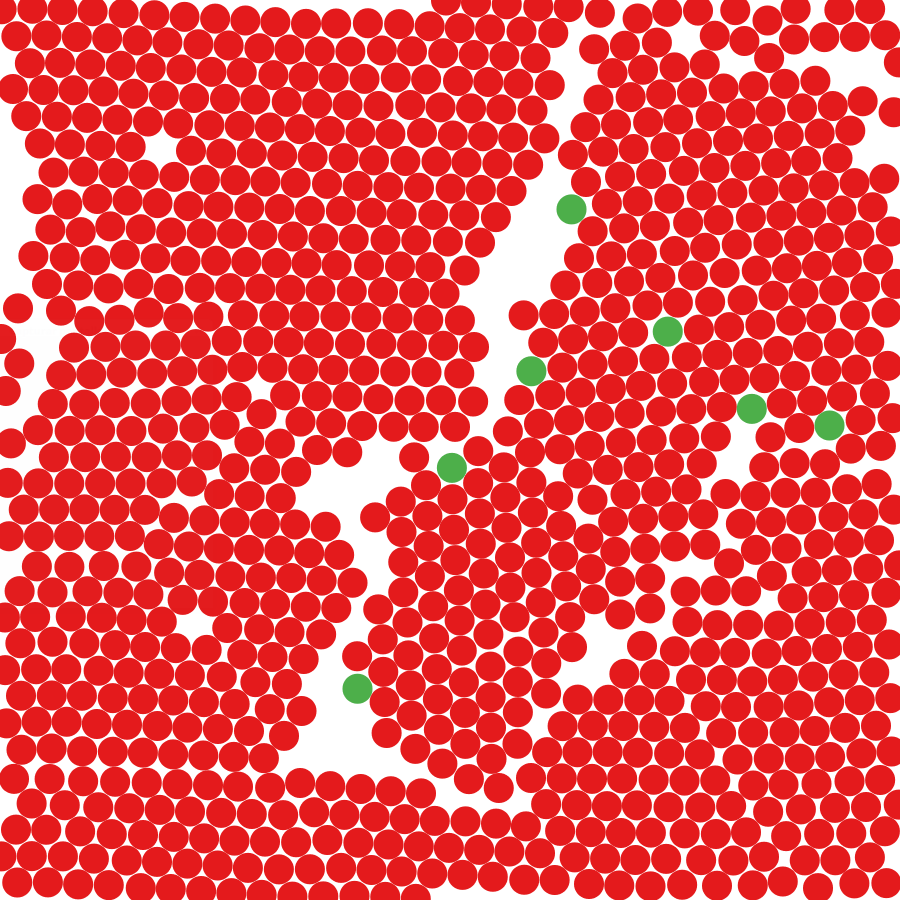}}
	\hfill
	\subcaptionbox{$\Delta t = 10^{-3}$}{\includegraphics[scale=0.12]{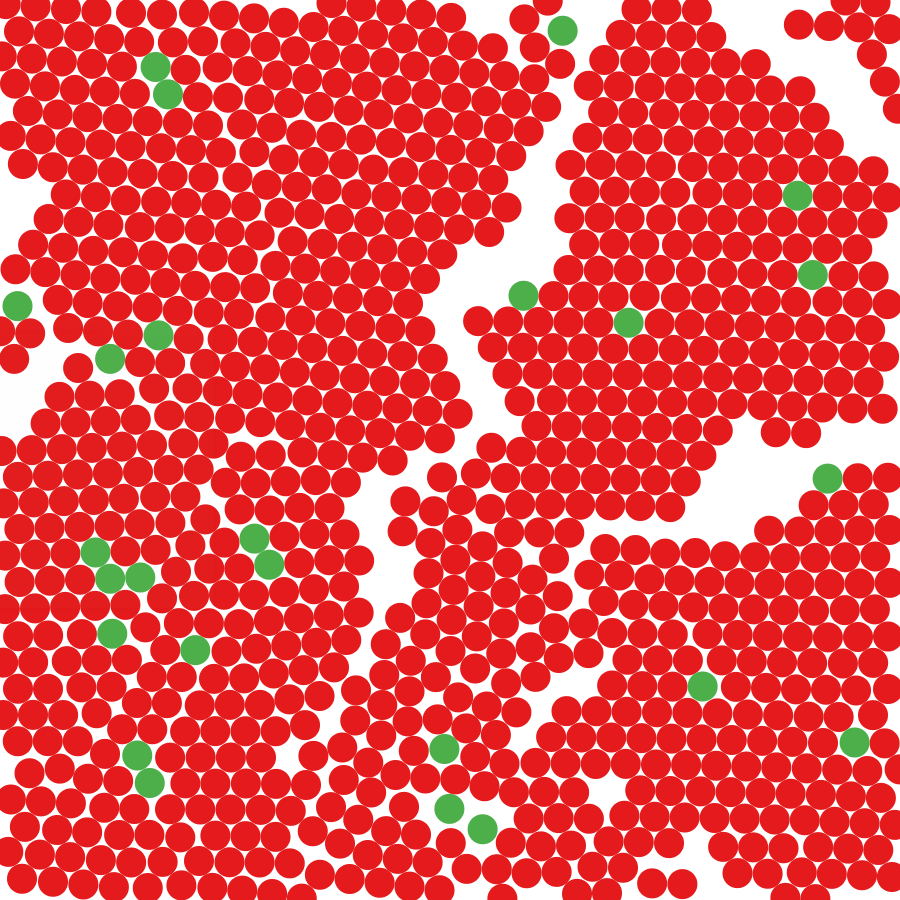}}
	\hfill
	\subcaptionbox{$\Delta t = 10^{-4}$}{\includegraphics[scale=0.12]{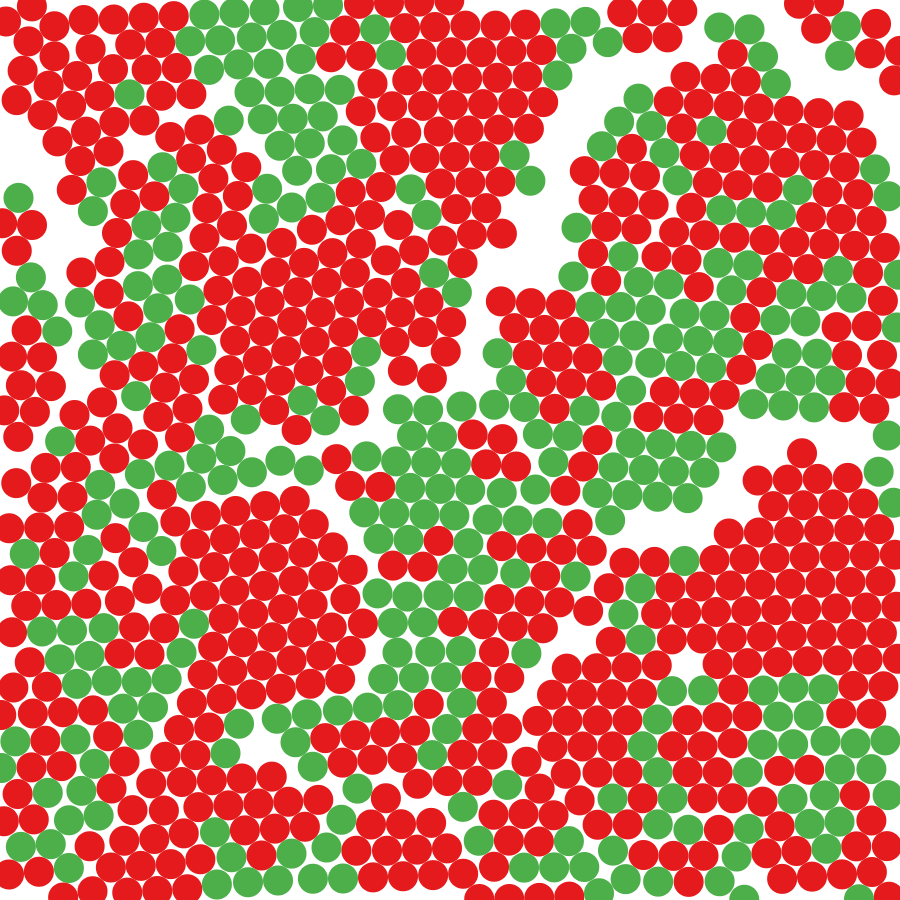}}	
	\caption{Overlap correction locality visualizations of configurations $c0$ and $c3$. The red particles participated in at least one overlap correction on the same time step, while green particles did not.}
	\label{fig:locality}
\end{figure}

In order to verify that the developed overlap correction algorithm is efficient enough, we made two validation experiments. First, we test the locality of the correction, that is, how far it propagates though the system. For configurations $c0$ and $c3$ we print the result after  $10^5$ time step iterations, painting with red the particles that took part in overlaps during the last simulated time step. Particles that did not take part in overlaps were painted green, so that every particle has a color. We repeat the process for decreasing values of $\Delta t$, expecting that the number of overlaps will decrease as the time step produces smaller movements. The results on Figure \ref{fig:locality} allows us to verify the complexity factor associated to each configuration that shows up on the previous performance curves. Configuration $c0$ involves much less particles on overlap corrections than $c3$ upon lowering the time step. This decrease in execution time by reducing $\Delta t$ does not compensate, however, for the larger number of steps that are needed to achieve an specified physical time. 

The second validation consists in testing the overlap correction on another colloidal model. We consider the Active Brownian Particle (ABP) model~\cite{ABP},  where particles move in 2D with velocities of fixed magnitude $V_0$, with a direction that is specified by the director angle $\theta_i$. The integration rule for the positions after an interval $\Delta t$ is:
\begin{equation}
\vec{r}_i(t + \Delta t) = \vec{r}_i(t) + V_0(\cos \theta_i \hat{x} + \sin \theta_i \hat{y}) \Delta t.
\end{equation}
In the same time interval, the angles $\theta_i$ are subjected to diffusive rotational Brownian motion, of amplitude $D$, and therefore evolve as:
\begin{equation}
\theta_i(t + \Delta t) = \theta_i(t) + \sqrt{2D \Delta t}\, n_i,
\end{equation}
where $n_i$ is a random Gaussian variable of zero mean and unit variance.

\begin{figure}[H] \centering
    \subcaptionbox{2500 iterations.}{\includegraphics[scale=0.17]{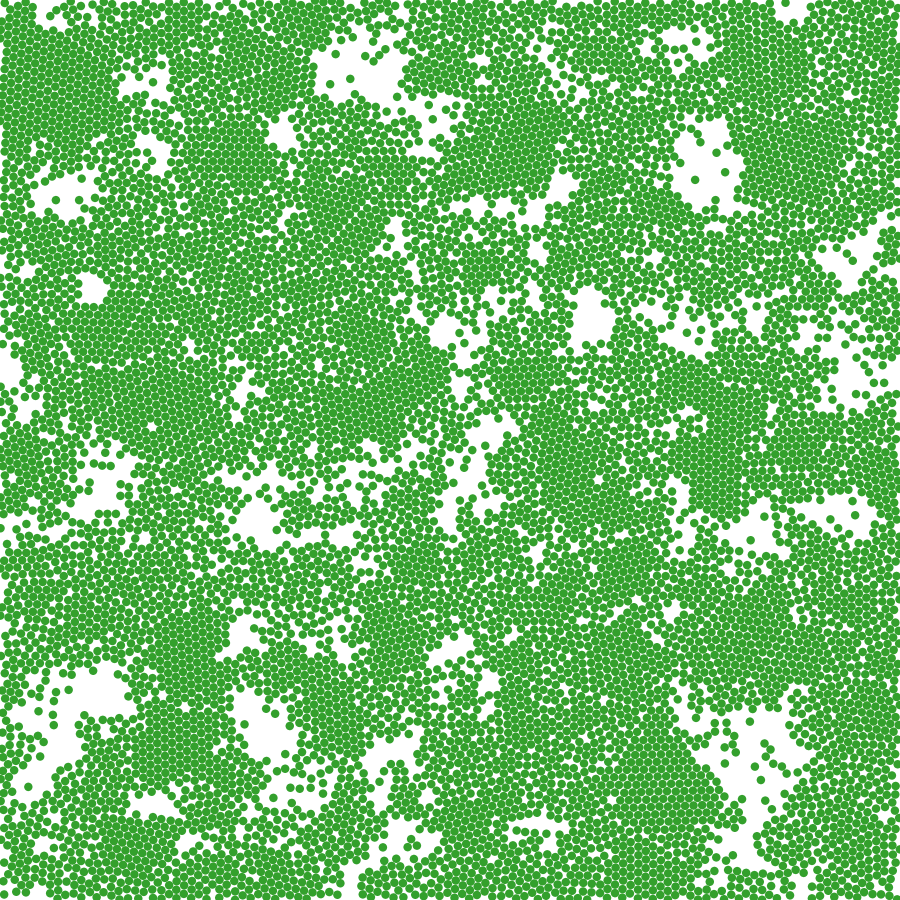}}
	\hfill
	\subcaptionbox{5000 iterations.}{\includegraphics[scale=0.17]{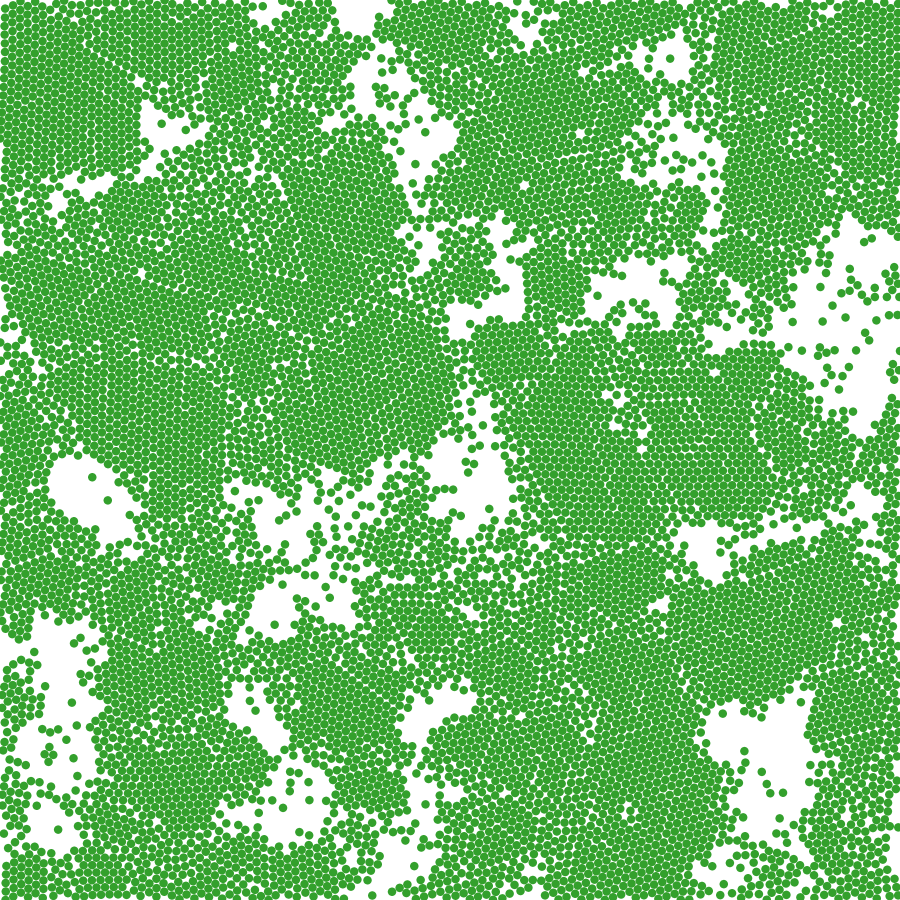}}
    \hspace{1cm}
	\subcaptionbox{7500 iterations.}{\includegraphics[scale=0.17]{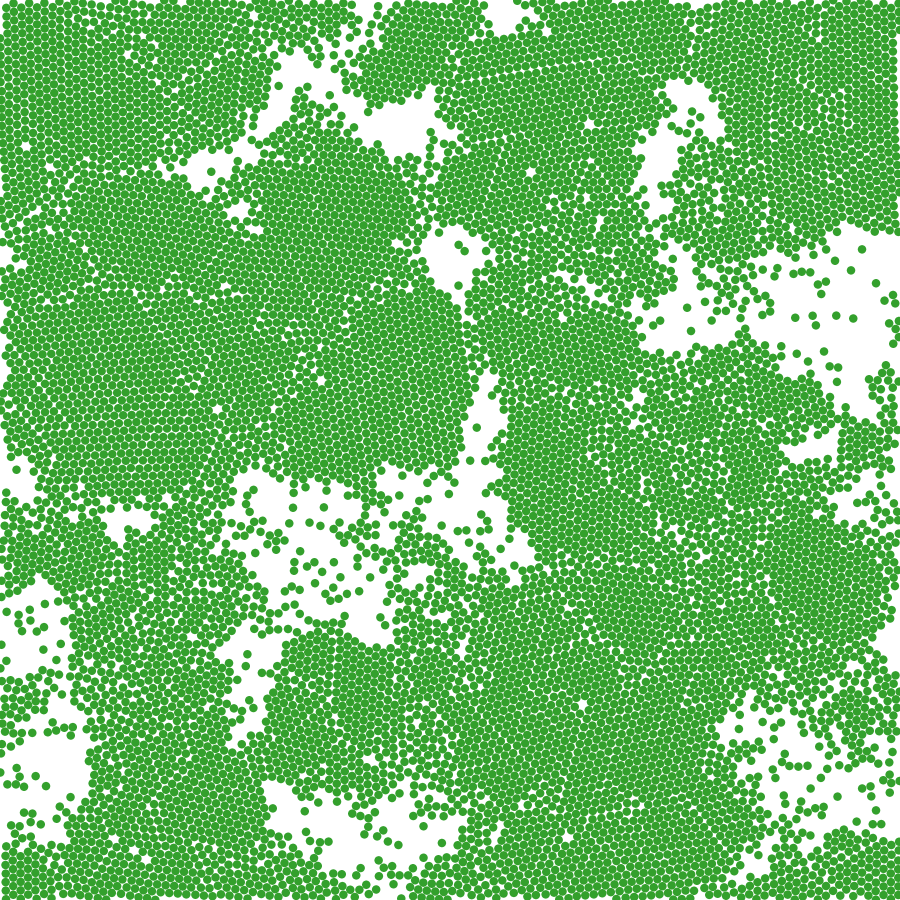}}
	\hfill
    \subcaptionbox{10000 iterations.}{\includegraphics[scale=0.17]{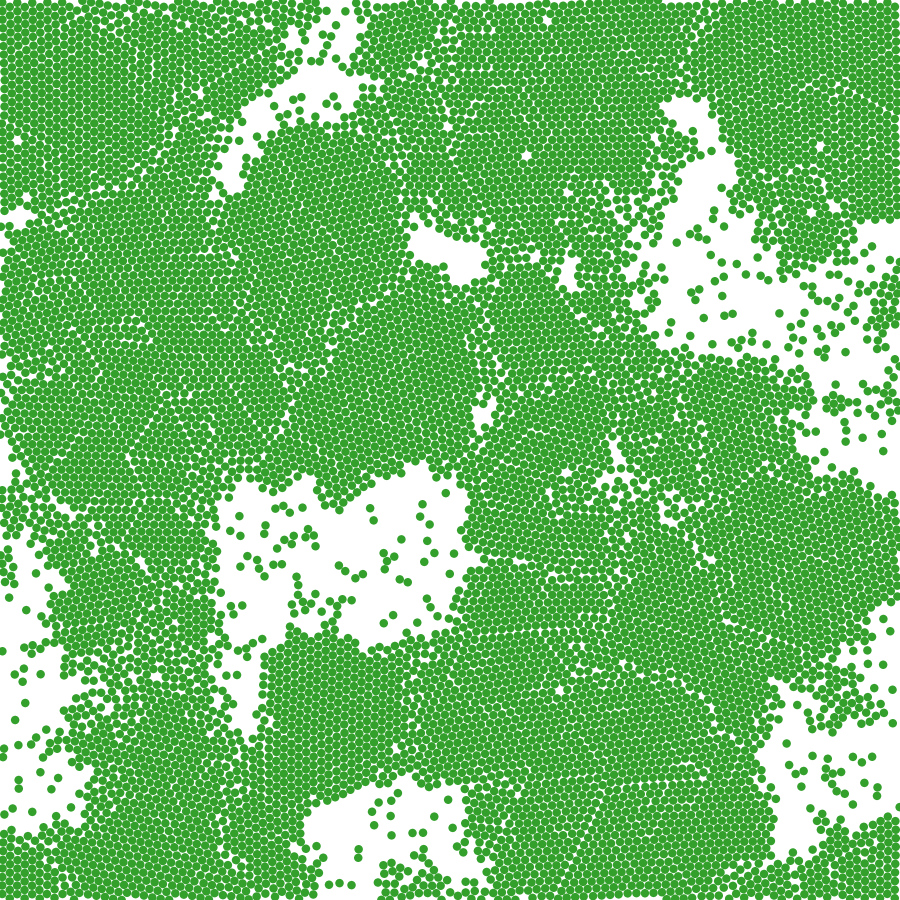}}
    
    \caption{Snapshots of an ABP system $N = 10^4, L = 105.9, D = 0.01$ and packing fraction $\rho = 0.7$.} 
	\label{fig:abp1}
\end{figure}

\begin{figure}[H] \centering
	\subcaptionbox{2500 iterations.}{\includegraphics[scale=0.17]{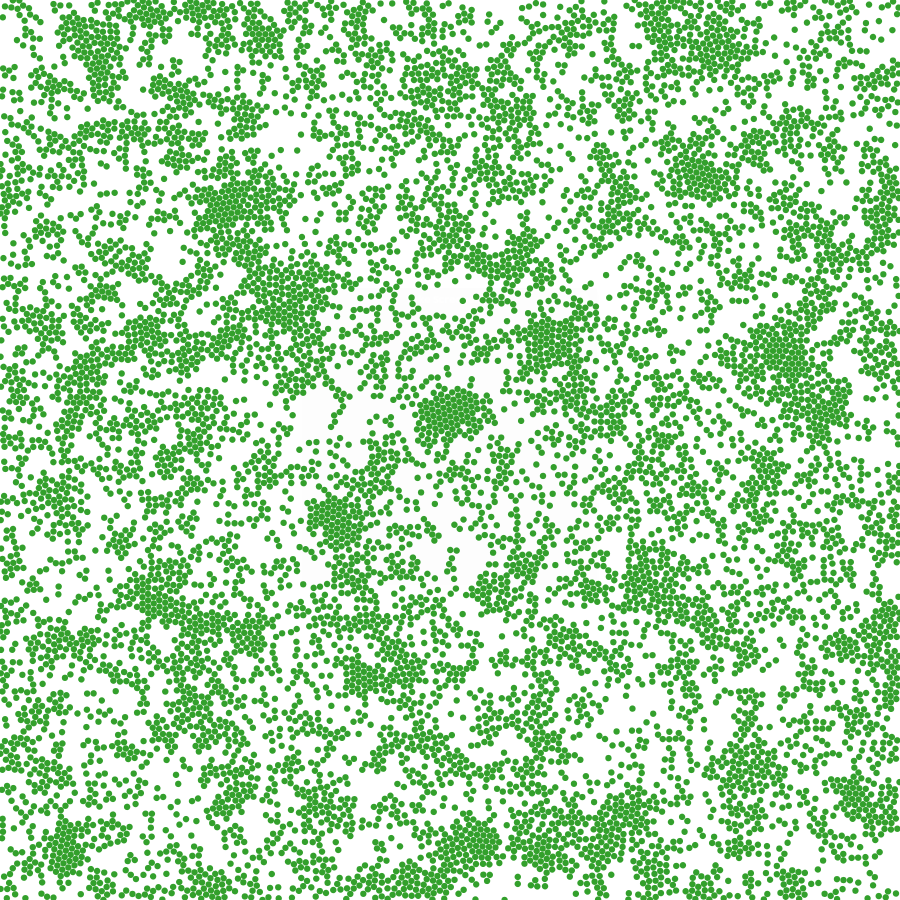}}
	\hfill
	\subcaptionbox{5000 iterations.}{\includegraphics[scale=0.17]{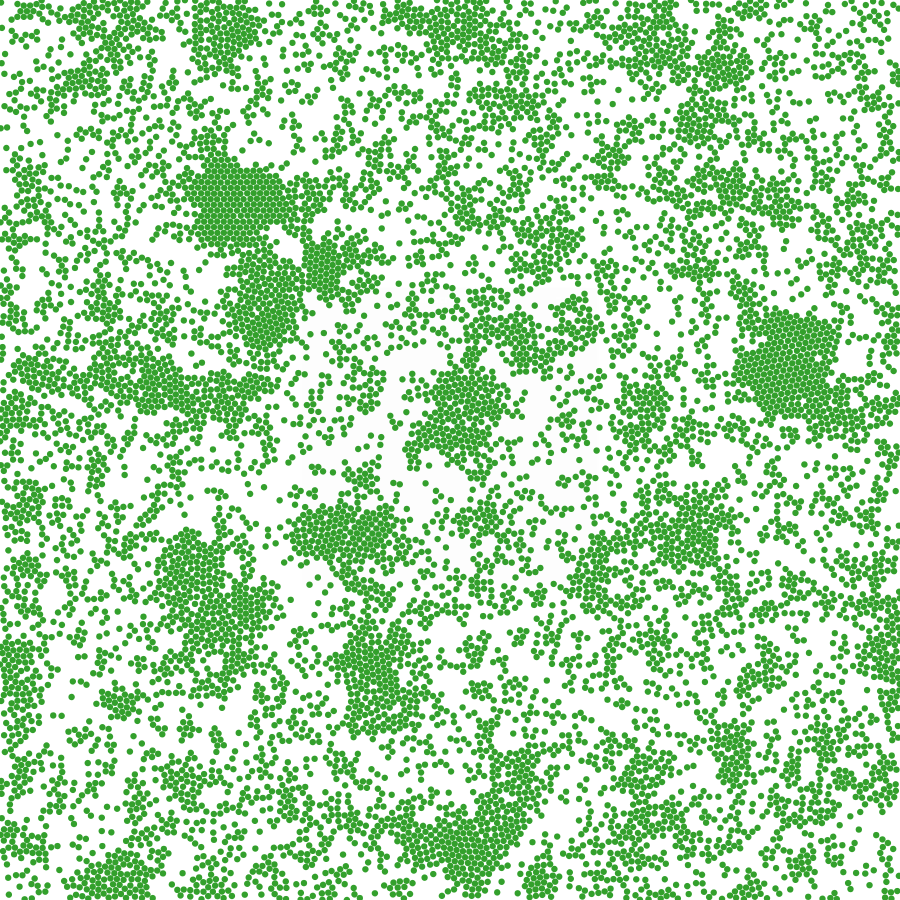}}
    \hspace{1cm}
	\subcaptionbox{7500 iterations.}{\includegraphics[scale=0.17]{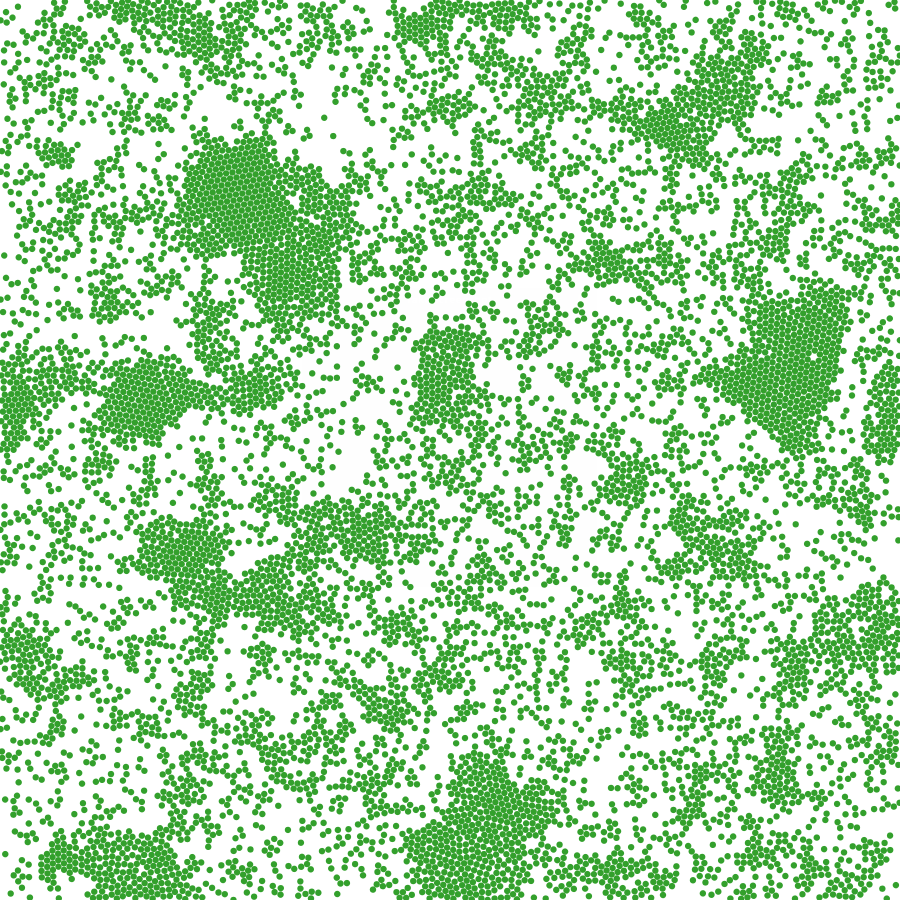}}
	\hfill
    \subcaptionbox{10000 iterations.}{\includegraphics[scale=0.17]{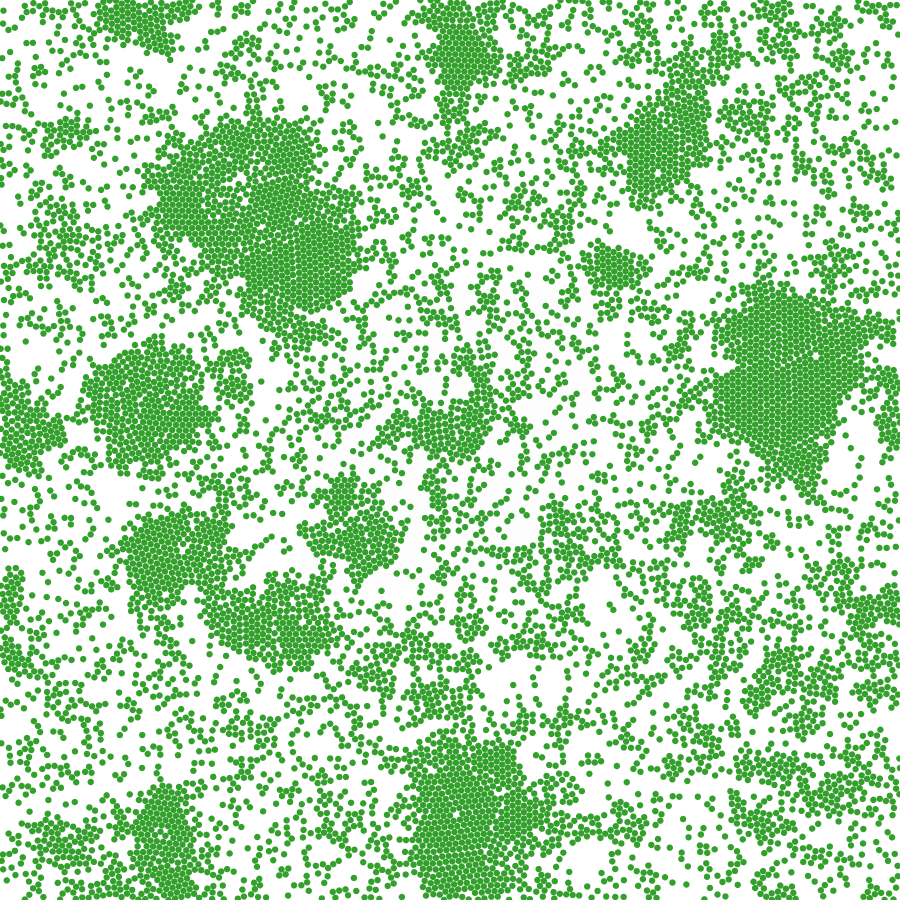}}
    
    \caption{Snapshots of an ABP system rescaled to fit the packing fraction $\rho = 0.4$. The other parameters are the same as presented on Figure \ref{fig:abp1}.}
	\label{fig:abp2}
\end{figure}

We simulate the system with the same parameters used in Ref.~\cite{Fily2012}, for two different packing fractions, obtaining the same phenomenlogy. At large packing fractions, the system evolves to the formation of a dense percolating dynamic cluster (see Fig.~\ref{fig:abp1}). Reducing the packing fraction, small clusters form, which merge in a slow coarsening process in the course of time as shown in Fig.~\ref{fig:abp2}.

\section{Discussion} \label{ch:discussion}

We presented algorithms for simulating colloidal particles subject to Brownian motion, interacting with short or long range force interactions, and presenting excluded volume. The overlap correction algorithm using Delaunay triangulations is a novel method. The algorithms implemented in CUDA for simulation are fully parallel, transferring data back to host only for measurements or outputs. The overlap correction algorithm can be used independently from the forces calculation, allowing to simulate different colloidal models including changed particles or self-propelled active systems. The Delaunay triangulation and the parallel edge-flip algorithm proved to be useful for solving overlaps efficiently. This opens the possibilty for using the Delaunay triangulation for solving related problems in the simulation, such as short range force calculation or approximated n-body simulations. The parallel n-body implementation was also successfully adapted and optimized to the particular conditions of colloidal particles, which opens up simulations of up to two orders of magnitude the number of particles used on the previous sequential implementation.

\section*{Acknowledgements}

The authors would like to thank the NVIDIA GPU Research Center of the Department of Computer Science of the Universidad de Chile for supplying the equipment used for the tests presented here. This work was partially supported by the FONDECYT projects No.~1140778 and No.~3160182, and by project No.~ENL009/15, VID, Universidad de Chile.

\end{document}